\documentclass[10pt, a4paper,natbib]{elsart}
\usepackage{amsfonts} 
\usepackage{graphicx}
\usepackage{latexsym}

\newtheorem{theorem}{Theorem}
\newtheorem{lemma}[theorem]{Lemma}

\newtheorem{definition}[theorem]{Definition}
\newtheorem{proposition}[theorem]{Proposition}

\newenvironment{proof}[1]{ \trivlist \item[\hskip \labelsep{\bf
#1}]}{\hfill\mbox{$\Box$} \endtrivlist} \newcommand{\junk}[1]{}

\begin{document}

\newcommand{\R}{{\Bbb R}} 
\newcommand{\D}{{\Bbb D}}
\newcommand{\C}{{\Bbb C}} 
\newcommand{\N}{{\Bbb N}} 
\newcommand{\Z}{{\Bbb Z}}
\newcommand{\Q}{{\Bbb Q}} 
\newcommand{\A}{{\Bbb A}}

\def\gcd{{\rm{gcd}\,}} \def\lc{{\rm{lc}\,}} \def\tc{{\rm{tc}\,}}
\def\cf{{\rm{cf}\,}} \def\val{{\rm{val}\,}} \def\Sylv{{\rm{Sylv}}}
\def\K{{\rm{K}}} \def\co{{\rm{co}}} \def\squo{{\rm{squo}\,}}
\def\srem{{\rm{srem}}} \def\rem{{\rm{rem}}} \def\quo{{\rm{quo}}}

\def\addots{\mathinner{\mkern1mu \raise1pt\vbox{\kern7pt\hbox{.}}
\mkern2mu\raise4pt\hbox{.}\mkern2mu \raise7pt\hbox{.}\mkern1mu}}

\begin{frontmatter}

\title{ Symmetric Subresultants and Applications }

\author{Cyril Brunie}
\ead{brunie@unilim.fr}
\address{Universit\'e de
Limoges, D\'epartement de Math\'ematiques, 126 av. Albert Thomas,
87060 Limoges C\'edex, France.}

\author{Philippe Saux Picart}
\ead{sauxpica@univ-brest.fr}
\address{Universit\'e de Bretagne Occidentale,  D\'epartement de
Math\'ematiques, 6 av. Victor Le Gorgeu, 29285 Brest C\'edex,
France.}

\begin{abstract}

\textsc{Schur}'s transforms of a polynomial are used
to count its roots in the unit disk. These are generalized them by introducing the
 sequence of symmetric sub-resultants of two
polynomials. Although they do have a determinantal definition, we show that
they satisfy a structure theorem which allows us to compute them with a
type of Euclidean division. As a consequence, a fast algorithm based on
a dichotomic process and FFT is designed.

We prove also that these symmetric sub-resultants have a deep link with
\textsc{Toeplitz} matrices. Finally, we propose a new algorithm of
inversion for such matrices. It has the same cost as those
already known, however it is fraction-free and consequently well adapted to
computer algebra.

\end{abstract}


\end{frontmatter}
\section{Introduction}

Let $P=a_{0}+a_{1}X+\cdots+a_{d}X^d$ be a polynomial in $\C[X]$.   In 1918 \textsc{Schur}
gave a method to compute the number of roots of $P$ in the
unit disk \cite{SC}. This work was completed by \textsc{Cohn} in 1922 \cite{CO}.

 The so-called \textsc{Schur-Cohn} algorithm works as follows.  Suppose
 that $a_{0}a_{d}\not=0$ and define the reciprocal of $P$ by $P^*=X^d
 \bar{P}(1/X)$.  Compute the following sequence of polynomials :
 $$T(P)=\overline{P(0)}P-\lc(P)P^*, \;\; T^k(P)=T(T^{k-1}(P)),$$

\noindent where $\lc(P)$ denotes the leading coefficient of $P$. This sequence
is finite : it has at most $\deg(P)$ polynomials with decreasing degrees and
real constant terms. It is the variation of the signs of these constant terms, all
supposed to be non-zero, which gives us the number of roots of $P$ in
the unit disk. See \textsc{Henrici} \cite{HEN} or \textsc{Marden} \cite{MA} for a precise description of this algorithm.

In this primary version, two difficulties arise.  First, the algorithm does not
work for every polynomial : if the difference of the degrees of two successive
transforms $T^k(P)$ is more than one, or if some constant terms are zero, it is
not possible to compute the number of roots of $P$. Second, the exact computation of these
transforms suffer from an exponential increase of the size of the coefficients :
at each step, the length of the coefficients is  approximately doubled.

For these two reasons, we introduced the new sequence of
\textsc{Schur-Cohn subtransforms} (see \textsc{Saux Picart}
\cite{SP}).  These subtransforms are equal to $T^k(P)$ up to a
multiplicative factor, can be computed for every polynomial, have a
determinantal definition, and an approximately linear increase is
their coefficients.  Moreover from the constant terms, we can compute
the number of roots of the polynomial in the unit disk, using an
adapted rule of signs.

Later on, it appeared that the sequence of the \textsc{Schur-Cohn subtransforms}
is linked to the sequence of the successive remainders of $P$ and $P^*$ in a
special ``symmetric" division (see \textsc{Brunie} and \textsc{Saux Picart}
\cite{BSP}).  This division consists in eliminating from the largest polynomial
as many monomials as possible from the top as well as from the tail by adding
good multiples of the ``divisor". In the article cited above, we give a structural
theorem, which describes the link between these two sequences built from $P$. 

In the present article we generalise the definition of the
\textsc{Schur-Cohn subtransforms} and the symmetric division of two
polynomials to a general situation (no restriction on $P$ and $P^*$).
We will speak of \textsc{symmetric subresultants } of two polynomials.
We are then able to formulate a new general ``structure-theorem" which
constitutes a central result of our work.  With this, we compute the
sequence of symmetric subresultants, using a \textsc{Euclid}-like
algorithm instead of the determinantal definition.  A dichotomic
process and DFT allow us to produce a fast algorithm.  Our methods are
adaptated from ideas introduced by \textsc{Sch\"{o}nage} for the
computation of Euclidean remainder sequences in \cite{SH}, and by
\textsc{Lickteig} and \textsc{Roy} in \cite{LR} for the computation of
classical subresultants.  The algorithm cost is of ${\mathcal{ O
}}({\mathcal{ M }}(d) \log d)$ arithmetical operations, where
${\mathcal{ M }}(d)$ denotes the cost of the multiplication of two
polynomials of degree $d$.

We will not describe the application to the number of roots of a
polynomial in the unit disk as it has already been discussed in
\cite{BSP}.  However there are well-known relations between the
problem of root isolation and \textsc{Toeplitz} matrices (see for
example, \textsc{M.G. Krein} and \textsc{M.A. Naimark} \cite{KN}).  We
use these links to give, in the last part, a fast algorithm for
solving \textsc{Toeplitz} systems with exact computation.  It has the
same cost as the well-known algorithm of \textsc{Brent},
\textsc{Gustavson} and \textsc{Yun} in \cite{BGY}, or those of
\textsc{Gemigniani} in \cite{GE1}.  Morover, it is fraction free and
consequently well adapted to computer algebra.  We also give a new way
to compute the signature of a Hermitian \textsc{Toeplitz} matrix.

This paper is organised as follows. Section 2  introduces  notations and 
definitions. In Section 3, we state the structure-theorem.
Section 4 describes how to efficiently compute the symmetric subresultants and 
the last section applies these results to \textsc{Toeplitz} matrices. 

Finally, we wish to  thank \textsc{M.-F. Roy}
and \textsc{T. Lickteig} for their help and interest in this work.

\section{Definitions and Notations}

Consider a subring $\D$ of $\C$ and define the valuation of a nonzero
polynomial $P\in \D[X]$, denoted by $v(P)$, as the greatest integer
$v$ such that $X^v$ divides $P$ (it is also named ''X-adic valuation``
in many books).  For the zero polynomial put $\deg (0)=-\infty$ and
$v(0)=\infty$.  Denote by $\D'$ the quotient field of $\D$.

We write $\co _{k}(P)$ for the coefficient of order $k$ of $P$.  If $\deg P=d$,
the leading coefficient $\co _{d}(P)$ is $\lc(P)$ and the trailing
coefficient $\co _{v(P)}(P)$ is denoted by $\tc(P)$. Remark : if $v(P)\not=0$,
$\tc(P)$ is different from $P(0)$.

We will use Euclidean division of a polynomial $A$ by a polynomial
$B$ in $\D[X]$~: the notation $\quo(A,B)$ stands for the quotient and
$\rem(A,B)$ for the remainder; they have their coefficients in the
fraction-field $\D'$.  We say that the division is \textit{exact} if
$\quo(A,B)$ and $\rem(A,B)$ are elements of $\D$.  Please note~: our
definition of exact division differs from another definition common in
the literature where exact division simply means vanishing of the
Euclidean remainder.

Now, let us introduce the main object of our article.

\subsection{Symmetric Subresultants}

Let $A=\sum_{i=0}^{d}a_i X^i$ and $B=\sum_{i=0}^{d}b_iX^i$ be two
polynomials in $\D[X]$.  We suppose that one of them at least, say
$A$, has its degree equal to $d$ ; $B$ can also be formally considered as
having degree $d$ : if $\deg B=d'<d$, $B$ will be replaced by $0X^d+
\cdots + 0X^{d-d'+1}+B$.  We also assume that the valuation is $0$ for
at least one of them, otherwise we divide both polynomials by a power
of $X$ to ensure this condition.  Define :

$$ \Sylv_j(A,B)= \underbrace{ \left( \matrix{
a_0&\cdots&\cdots&\cdots&\cdots&a_d&& \cr
&\ddots&\ddots&\ddots&\ddots&\ddots&\ddots& \cr
&&a_0&\cdots&\cdots&\cdots&\cdots&a_d\cr b_0&\cdots&\cdots&\cdots&\cdots&b_{d}&&
\cr &\ddots&\ddots&\ddots&\ddots&\ddots&\ddots& \cr
&&b_0&\cdots&\cdots&\cdots&\cdots&b_{d} \cr } \right) }_{d+j} \matrix{
\left.\matrix{\cr \cr \cr}\right\}&{j} \cr \left.\matrix{\cr \cr \cr
}\right\}&{j} \cr} $$ 

to be a submatrix of the full \textsc{Sylvester} matrix
$\Sylv_d(A,B)$.\\ For $\ell=0,\ldots,d-j$, let
$\Sylv_{j,\ell}=\Sylv_{j,\ell}(A,B)$ be the following $2j \times 2j$ square
submatrix of $\Sylv_j(A,B)$  :

$$ \Sylv_{j,\ell}=\underbrace{\left( \matrix{ a_0 & \cdots & a_{j-2} \cr &\ddots
&\vdots \cr & & a_0  \cr & & 0 \cr b_0&\cdots&b_{j-2}\cr & \ddots & \vdots \cr &
& b_{0}  \cr & & 0 \cr } \right. }_{j-1} {\left. \matrix{ a_{j-1+\ell} \cr
\vdots \cr a_{\ell+1} \cr a_{\ell} \cr b_{j-1+\ell} \cr \vdots \cr b_{\ell +1}
\cr b_{\ell} \cr } \right.} \underbrace { \left. \matrix {a_d & & &\cr
\vdots&\ddots & & \cr a_{d-j+2} & \cdots&a_d & \cr a_{d-j+1} & \cdots&a_{d-1} &
a_d \cr b_d & & &\cr \vdots& \ddots& &\cr b_{d-j+2} & \cdots & b_{d} & \cr
b_{d-j+1} & \cdots&b_{d-1} & b_d \cr} \right)}_{j} \matrix{ \left.\matrix{\cr
\cr \cr \cr \cr}\right\}&{j} \cr \left.\matrix{\cr \cr \cr \cr \cr }\right\}&{j}
\cr}.$$

The sequence $(S_{j})_{-1 \le j \le d}$ of  symmetric
subresultants of $A$ and $B$ is  defined by~:

\medskip \hspace{1cm} \begin{minipage}{10cm} {\begin{itemize} \item $S_{-1}= A$,
\item $S_0 = B,$ \item $S_{j}(A,B)= \sum_{\ell=0}^{d-j}
\det(\Sylv_{j,\ell})X^\ell$, \quad if $1 \leq j \leq d $.
\end{itemize}} \end{minipage}

\medskip Clearly,  $S_{j}$ is an element of $\D[X]$ for any $j$.
The last one, $S_{d}$ is just the resultant of $A$ and $B$.  In the generic
situation, $S_{j}$ is of degree $d-j$ and valuation 0. However, the real degree
could be less than $d-j$ and the valuation greater than 0. In order to describe these
situations, we introduce the following definition. 

Let $(\alpha,\beta)$  be such that :
 $$ \left\{ \begin{array}{ccc} v(S_{j}) & = & 0
\\ \deg(S_{j}) & = & d-j \end{array} \right.  \;\;\mathrm{and}\;\;
\left\{ \begin{array}{ccc} v(S_{j+1}) & = & \alpha  \\ \deg(S_{j+1}) & =
& d-j-\beta \end{array} \right. ,  $$

we will then say that the pair $(S_{j},S_{j+1})$ is $\mathbf{(\alpha,
\beta)}$\textbf{-defective}. The case $(0,1)$ is just the general situation
without special deflation.

Just as for the classical subresultants, we can express the $S_{j}$ through a
\textsc{Bezout} relation between $A$ and $B$. This is estblished in the next
lemma.

\begin{lemma}\label{relbezout} Let $A$ and $B$ be two polynomials in $\D[X]$ of
the same degree $d$ and valuation 0. For every $j \in \{0, 1,\ldots,d-1 \},$
there exist two elements in $\D[X]$, $U_{j}$ and $V_{j}$, such that :
$$X^{j}S_{j+1}= U_{j}A+ V_{j}B.$$ 
The degrees of $U_j$ and $V_j$ are at
most $j$. These polynomials are unique under such an assumption.
\end{lemma}

\begin{proof}{Proof : } Using a matrix with coefficients in $\D[X]$, we can
write $X^{j}S_{j+1}$ as  a determinant in the following way :
    
    $$ X^{j}S_{j+1}=\left | \begin{array}{ccccccc}
a_{0} & \ldots & a_{j-1} & X^{j}a_{j}+\ldots+X^{d-1}a_{d-1} & a_{d} &  &
\\

& \ddots & \vdots & X^{j}a_{j-1}+\ldots+X^{d-1}a_{d-2}  & \vdots &
\ddots &   \\

&  &  & \vdots &  &  &   \\

&  & a_{0} & X^{j}a_{1}+\ldots+X^{d-1}a_{d-j+1} & \vdots &  &   \\

&  &  & X^{j}a_{0}+\ldots+X^{d-1}a_{d-j} & a_{d-j+1}  & \ldots  &
a_{d}\\

\rule{0pt}{12pt} b_{0} & \ldots & b_{j-1} &
X^{j}b_{j}+\ldots+X^{d-1}b_{d-1} & b_{d} &  &   \\

& \ddots & \vdots & X^{j}b_{j-1}+\ldots+X^{d-1}b_{d-2}  & \vdots &
\ddots &   \\

&  &  & \vdots &  &  &   \\

&  & b_{0} & X^{j}b_{1}+\ldots+X^{d-1}b_{d-j+1} & \vdots &  &   \\

&  &  & X^{j}b_{0}+\ldots+X^{d-1}b_{d-j} & b_{d-j+1}  & \ldots  &
b_{d}\\

    \end{array} \right | .  $$
    
    We do not change the value of this determinant by adding to the $( j+1 )$-th
    column a linear combination of the other ones. More precisely, call \(
    C_{i} \) the \( i \)-th column of the above matrix ($ i= 1,\ldots,
    2j+2$). Then add to the $( j+1)$-th column \( C_{1}+XC_{2}+\ldots +
    X^{j-1}C_{j}+X^dC_{j+2}+\ldots+X^{d+j}C_{2j+2} \). We obtain :

    $$ X^{j}S_{j+1}= \left | \begin{array}{ccccccc}
a_{0} & \ldots & a_{j-1} & A & a_{d} &  &   \\

& \ddots & \vdots & XA  & \vdots & \ddots &   \\

&  &  & \vdots &  &  &   \\

&  & a_{0} & X^{j-1}A & \vdots &  &   \\

& & &X^{j}A&

a_{d-j} & \ldots & a_{d}  \\
b_{0} & \ldots & b_{j-1} & B & b_{d} &  &   \\

& \ddots & \vdots & XB  & \vdots & \ddots &   \\

&  &  & \vdots &  &  &   \\

&  & b_{0} & X^{j-1}B & \vdots &  &   \\
         
&  &  & X^{j}B &

b_{d-j} & \ldots & b_{d}  \\

    \end{array} \right | .  $$
    
Expand this determinant according to the $(j+1)$-th column, putting $A$
as a factor in the  first $j+1$ lines and $B$ in the last $j+1$ : therefore
 there exist two polynomials, $U_{j}$ and $V_{j}$, of degree at most
$j$ such that : $$X^{j}S_{j+1}= U_{j}A+V_{j}B.$$ Furthermore, we can express
these polynomials as determinants. We have :

$$ U_{j}= \left | \begin{array}{ccccccc}
a_{0} & \ldots & a_{j-1} & 1 & a_{d} &  &   \\

& \ddots & \vdots & X  & \vdots & \ddots &   \\

&  &  & \vdots &  &  &   \\

&  & a_{0} & X^{j-1} & \vdots &  &   \\

& & &X^{j}&

a_{d-j} & \ldots & a_{d}  \\
b_{0} & \ldots & b_{j-1} & 0 & b_{d} &  &   \\

& \ddots & \vdots & 0  & \vdots & \ddots &   \\

&  &  & \vdots &  &  &   \\

&  & b_{0} & 0 & \vdots &  &   \\
         
&  &  & 0 &

b_{d-j} & \ldots & b_{d}  \\

    \end{array} \right | , $$ and :

$$ V_{j}= \left | \begin{array}{ccccccc}
a_{0} & \ldots & a_{j-1} & 0 & a_{d} &  &   \\

& \ddots & \vdots & 0  & \vdots & \ddots &   \\

&  &  & \vdots &  &  &   \\

&  & a_{0} & 0 & \vdots &  &   \\

& & &0&

a_{d-j} & \ldots & a_{d}  \\
b_{0} & \ldots & b_{j-1} & 1 & b_{d} &  &   \\

& \ddots & \vdots & X  & \vdots & \ddots &   \\

&  &  & \vdots &  &  &   \\

&  & b_{0} & X^{j-1} & \vdots &  &   \\
         
&  &  & X^{j} &

b_{d-j} & \ldots & b_{d}  \\

    \end{array} \right |.  $$ \end{proof}

For $j=0$, we simply have $S_{1}=b_{d}A-a_{d}B$, \textit{i.e.} $U_{0}=b_{d}$ and
$V_{0}=-a_{d}$. Using the determinantal definition of $U_{j}$ and $V_{j}$, we
see that : $$U_{j}(0)=b_{0}\cdot \co _{d-j}(S_{j}), \;\;\; \co
_{j}(U_{j})=\lc(U_j)=b_{d}\cdot S_{j}(0),$$ and also :
$$V_{j}(0)=-a_{0}\cdot  \co _{d-j}(S_{j}), \;\;\; \co _{j}(V_{j})=\lc(V_j)=-a_{d}\cdot
S_{j}(0).$$ 
Finally, we can observe that these polynomials are uniquely 
determined, first when $A$ and $B$ are co-prime, and then in the 
general case.  (The proof uses the same arguments as for the
extended Euclidean algorithm for polynomials; see \cite{GCL}.)


\subsection{Symmetric division of polynomials}

The division we use is justified by the following lemma.

\begin{lemma} Let $A, B \in \D[X]$, with $B\neq 0$, $\deg A=d\ge \deg B=d-\beta$
and $v(B)=\alpha$.  There exist $Q,R \in \D'[X]$, where $\D'$ is the
fraction field of $\D$,  uniquely determined, such that $\deg Q=\alpha +
\beta$ and $\deg R < d-(\alpha+\beta)$, and : $$A=Q
\frac{B}{X^\alpha} + X^\beta R.$$

\end{lemma} 

\begin{proof}{Proof :}

We sketch how to compute $Q$ and $R$.  First divide $A$ by
$B/{X^\alpha}$ with increasing powers of $X$ up to order
$\beta$.  We obtain : 
$$A=Q_1 \frac{B}{X^\alpha} + X^\beta R_1,$$
with $\deg Q_1<\beta$ and $\deg R_1 = d-\beta$. Then, compute the
Euclidean division of $R_1$ by $B/{X^\alpha}$ : $$R_1= Q_2  \frac{B}{X^\alpha}
+R,$$

where $\deg R < d-\beta -\alpha$ and $\deg Q_2 = \alpha$. Then, define $Q$ by
$Q=Q_1+X^\beta Q_2$ to establish the claim. Uniqueness is proven as usual.
\end{proof}

The polynomial $Q$ is called the \textit{symmetric quotient} of $A$ by $B$, noted
$\squo(A,B)$ and $R$ the \textit{symmetric remainder}, denoted $\srem(A,B)$.

It is clear that the computation of such a division has the same arithmetical
cost as ordinary Euclidean division.  It requires, at most, 
$d(\alpha+\beta+1)$ arithmetical operations.

Historical note : We can find various kinds of ``symmetric'' division
introduced by authors with specific aims. See for exemple, \textsc{Jezek}
\cite{JJ}, \textsc{Demeure} and \textsc{Mullis} \cite{DM}. However, our
definition  is different from the one in \cite{JJ} and, when $\alpha=\beta$, coincides with the one given by \textsc{Demeure} and \textsc{Mullis} only in the case.



\section{Structure-Theorem for symmetric subresultants} \label{grostheo} We
 now describe the relationship between the sequence of symmetric subresultants
and the sequence of symmetric remainders of two polynomials. Our
main result is :

\begin{theorem}\label{bigtheo} Let $\D$ be a subring of $\C$, and let
$A$ and $B$ be elements of $\D[X]$ of degree $d$ and valuation 0.  Let
$(S_{i})_{0 \le i\le d}$
      be the sequence of symmetric subresultants of $A$ and $B$.  Suppose
      that the pair $(S_{j}, S_{j+1})$ is $(\alpha, \beta)$-defective. We have
      :

\begin{enumerate} \item    \begin{itemize} \item if $\alpha >0$ and $\beta>1$,
then $S_{j+k} \equiv 0$ for $k=2, \ldots ,\alpha+\beta-1$ 
      
      \item if $\alpha=0$ and $\beta>1$, then, if $j>0$ : $$ S_j(0) \cdot
S_{j+k}=S_{j+1}(0)^{k-1} S_{j+1} \;\;\mathrm{for} \;\;
k=2,\ldots, \beta-1.$$
      
      If $j=0$, $S_{k}=S_{1}(0)^{k-1} S_{1}$ for $k=2, \ldots , \beta-1$.
      
      \item if $\alpha>0$ and $\beta=1$, then if $j>1$ :
       $$ \lc(S_{j})^{k-1} \cdot S_{j+k}= (-1)^{k}\lc(S_{j+1})^{k-1} \cdot
      \frac{S_{j+1}}{X^{k-1}}\;\; \mathrm{for} \;\; k=2,\ldots ,\alpha.$$
       If $j=0$, $b_{d}^k \cdot S_{k}=(-1)^k \cdot \lc(S_{1})^{k-1}\ldots
      S_{1}X^{-k+1}$ for $ k=2,\cdots,\alpha$.  
      \end{itemize}
     
   \item In all cases, if $j>0$, we have :
   $$ \lc(S_j)^{\alpha} \cdot S_j(0)^{\beta-1} \cdot S_{j+\alpha+\beta} =
   (-1)^{(\alpha+\beta)\alpha} \cdot \lc(S_{j+1})^{\alpha}
   \cdot\tc(S_{j+1})^{\beta-1} \cdot {S_{j+1} \over X^\alpha },$$
   
   and if $j=0$, then : $$b_{d}^\alpha \cdot S_{\alpha+\beta} =
   (-1)^{(\alpha+\beta)\alpha} \cdot b_{0}^{\alpha} \cdot\lc(S_{1})^{\alpha}
   \cdot  \tc(S_{1})^{\beta-1} \cdot {S_{1} \over X^\alpha }.  $$
   
   \item In all cases, if $j>0$, we have :
   $$ \begin{array}{ccc} \lc(S_j) \cdot S_j(0) \cdot S_{j+\alpha+\beta+1} & = &
-\lc(S_{j+1}) \cdot S_{j+\alpha+\beta}(0) \cdot \srem(S_{j},S_{j+1})
\\ & = & -\srem\left(\lc(S_{j+1})\cdot S_{j+\alpha+\beta}(0)\cdot
S_{j},S_{j+1}\right) \end{array} $$
   
   and if $j=0$ then : $$b_{d}\cdot S_{\alpha+\beta+1} =
   -\srem\left(\lc(S_{1})\cdot S_{\alpha+\beta}(0)\cdot S_{0},S_{1}\right).$$
   
\end{enumerate}

\end{theorem} One remarkable fact is that the last symmetric divisions are
exact in $\D$, as we shall  prove later.

Observe that $S_1$ can also be expressed as a symmetric
remainder : by Lemma 1, we have $S_1=b_dA-a_dB=\srem(S_{-1},S_{0})$.

It could be helpful to the reader to visualize the different situations. 

\begin{enumerate} \item  Case $(S_{j},S_{j+1})$  defective on ``each side",
$\alpha > 0, \beta >1$ :
    
  $$  \begin{array}{lcl} & \hspace{0.5cm} & \hspace{3.2cm} \vdots \\ S_{j-1} &
\hspace{0.5cm} & \rule{6.5cm}{0.5mm}  \\ S_{j} & \hspace{0.5cm} &
\rule{6cm}{0.5mm}  \\ S_{j+1} & \hspace{0.5cm} &
\hspace{1cm}\rule{3.5cm}{0.5mm}  \\ &  &   \\ & &   \hspace{1.5cm}
\mathrm{Nullity}  \\ &  &   \\ S_{j+\alpha+\beta} & \hspace{0.5cm} &
\rule{3.5cm}{0.5mm}  \\ S_{j+\alpha+\beta+1} & \hspace{0.5cm}
&\rule{3cm}{0.5mm}   \\ & \hspace{0.5cm} & \hspace{1.5cm} \vdots
  \end{array} $$

    \item Case $(S_{j},S_{j+1})$  defective on the ``right-hand side", $\alpha = 0,
\beta >1$ : 
    
$$    \begin{array}{lcl} & \hspace{0.5cm} & \hspace{3.2cm} \vdots \\ S_{j-1} &
\hspace{0.5cm} & \rule{6.5cm}{0.5mm}  \\ S_{j} & \hspace{0.5cm} &
\rule{6cm}{0.5mm}  \\ S_{j+1} & \hspace{0.5cm} & \rule{3.5cm}{0.5mm}  \\
&  &  \hspace{1.7cm} $\vdots$ \\ &  &  \D-\mathrm{Proportionality}    \\ &  &
\hspace{1.7cm} \vdots \\
S_{j+\alpha+\beta} & \hspace{0.5cm} & \rule{3.5cm}{0.5mm}  \\
S_{j+\alpha+\beta+1} & \hspace{0.5cm} &\rule{3cm}{0.5mm}   \\ &
\hspace{0.5cm} & \hspace{1.5cm} \vdots \end{array} $$

    \item Case $(S_{j},S_{j+1})$  defective on the ``left-hand side", $\alpha >0,
\beta =1$ : 
    
$$    \begin{array}{lcl} & \hspace{0.5cm} & \hspace{3.2cm} \vdots \\ S_{j-1} &
\hspace{0.5cm} & \rule{6.5cm}{0.5mm}  \\ S_{j} & \hspace{0.5cm} &
\rule{6cm}{0.5mm}  \\ S_{j+1} & \hspace{0.5cm} &
\hspace{2cm}\rule{3.5cm}{0.5mm}  \\ &  &
\hspace{1.5cm}\rule{3.5cm}{0.5mm} \\ &  & \hspace{0.5cm}
\D[X]-\mathrm{Proportionality } \\ &  &
\hspace{0.5cm}\rule{3.5cm}{0.5mm}\\ S_{j+\alpha+\beta} & \hspace{0.5cm}
& \rule{3.5cm}{0.5mm}  \\ S_{j+\alpha+\beta+1} & \hspace{0.5cm}
&\rule{3cm}{0.5mm}   \\ & \hspace{0.5cm} & \hspace{1.5cm} \vdots
\end{array} $$ \end{enumerate}

\begin{proof}{Proof :}

Roughly speaking, we can say that the rows of $Sylv_{i}(A, B)$ are made of $A,
XA$,..., $ X^{i-1}A$, and  $B, XB,... , X^{i-1}B$, identifying the vectors of the
coefficients of these polynomials with the polynomials themselves. Furthermore,
we consider them all of formal degree $d+i-1$.

\textbf{\underline {Preliminary work}  :} By Lemma \ref{relbezout}, we know the
existence of two polynomials, $U_{j}=\sum_{i=0}^{j} u_{i}X^{i }$ and
$V_{j}=\sum_{i=0}^{j}v_{i}X^{i}$, such that :

\begin{eqnarray*} X^{j}S_{j+1} & =& U_{j}A+ V_{j}B \\ & = & \sum_{n=0}^{j} u_{n}
(AX^{n}) + \sum_{n=0}^{j} v_{n}(BX^{n}).  \end{eqnarray*}

As the pair $(S_{j}, S_{j+1})$ is $(\alpha, \beta)$-defective, $S_{j}(0)$ and
$\co_{d-j}(S_{j})= \lc (S_{j})$ are different from zero. Because of the
determinantal definition of  $U_j$ (see proof of Lemma \ref{relbezout}), we have : 

\hspace{1cm} -- if $j>0$, $u_{0}=b_{0}\cdot\lc(S_{j}) \not = 0$, and
$u_{j}=b_{d}\cdot S_{j}(0)\not = 0$,

\hspace{1cm} -- if $j=0$, $u_{0}=u_{j}=b_{d} \not =0$.

Then,  for every $\ell \ge 0$, we have : $$ X^{j+\ell} S_{j+1}=   \sum_{n=0}^{j}
u_{n} AX^{n + \ell} + \sum_{n=0}^{j} v_{n}BX^{n + \ell}, \;\;\; (\dag)$$ 

with $u_{0}$ and $u_{j}$ different from 0. 

For $k \ge 2$, and $i$ fixed between 0 and $k-1$, we can replace the
$(i+1)$-th row of $\Sylv_{j+k}$, $X^{i}A$ by the linear combination of
the rows $X^{i}A, ...  ,X^{j+i}A$ and $X^{i}B, ...  ,X^{j+i}B$
described in $(\dag)$.  For $\ell=i$ we obtain $X^{j+i}S_{j+1}$ on
the $(i+1)$-th row of $\Sylv_{j+k}$ instead of $X^{i}A$.  The minors
of order $2(j+k)$ of this new matrix are equal to $u_{0}$ times the
corresponding ones in $\Sylv_{j+k}$.  This operation will be called
the $(i, \downarrow)$-\textbf{transformation} of $\Sylv_{j+k}$.  The
downward arrow means that the $j$ rows directly below the $(i+1)$-st
row are used.

We define also the $(j+i, \uparrow)$-\textbf{transformation} for $i=0, ... ,
k-1$ : this replaces the $(j+i+1)$-st row by $X^{j+i}S_{j+1}$ which is a linear
combination of the rows $X^{i}A, ... ,X^{j+i}A$ and $X^{i}B, ... ,X^{j+i}B$, by
$(\dag)$. In this case the values of the minors of order $2(j+k)$ of
$\Sylv_{j+k}$ are multiplied by $u_{j}$.

We use these two transformations in four different situations, described
below. For each, we have drawn the corresponding matrix resulting from $S_{j+k}$:
 on each diagram, the
rows with large dash patterns delimit the $j+k-1$ first columns and the $j+k$ last
ones needed for the computation of $\Sylv_{j+k, \ell}$ ($\ell=0,... ,d-j-k$). The
shadowed triangles highlight the coefficients of the matrix needed for the
computation of $S_{j}(0)$.

We consider now the four different cases.

\begin{itemize}

\item $1 \le \beta \le \alpha$. Two situations have to be distinguished.

\begin{itemize}

\item[$\Diamond$] If $2 \le k \le \alpha$, we use $k$ $(i,
\downarrow)$-transformations for $i=0,... , k-1$ in this
order. We obtain the matrix $\mathbf{M_{1}}$ (fig.
                   1).

\begin{figure}[htb] \begin{center} \includegraphics{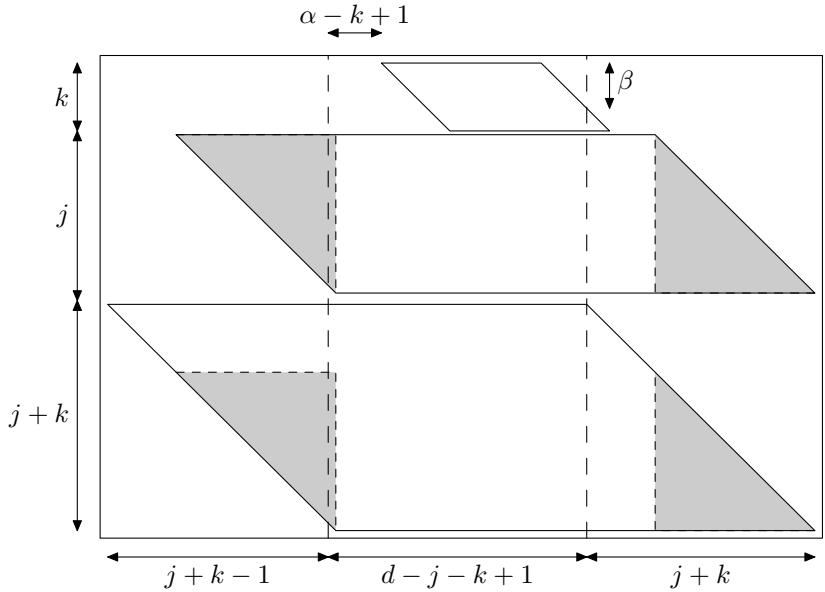} \end{center}
\caption{Shape of the matrix $M_{1}$} \end{figure}

\medskip
For each $\ell \in \{0,... , d-j-k\}$, the minor
$\det(\Sylv_{j+k, \ell})$ of $\Sylv_{j+k}$ is equal to
the corresponding minor of the above matrix divided by
$u_{0}^k$. If we denote this minor by $d_{j+k,\ell}$, we have : $$
u_{0}^k \cdot \det(\Sylv_{j+k,\ell})=d_{j+k, \ell}. $$

\item[$\Diamond$] If $\alpha < k \le \alpha + \beta$, we use $\alpha$ $(i,
\downarrow)$-transformations for $i=0,... , \alpha-1,$ in this
order, and then $k-\alpha$ $(j+i, \uparrow)$-transformations
for $i=k-1,... , \alpha,$ again in this order. We obtain the
matrix $\mathbf{M_{2}}$ (fig. 2). 
                   
\begin{figure}[htb]
\begin{center} \includegraphics{BSP.3} \end{center}
\caption{Shape of the matrix $M_{2}$}
\end{figure}

With the same notation as in the first case, we have : $$
u_{0}^\alpha \cdot u_{j}^{k-\alpha} \cdot
\det(\Sylv_{j+k,\ell}) = d_{j+k, \ell} . $$

\end{itemize}

\item $0 \le \alpha < \beta$.  Once again two situations occur.
\begin{itemize}

\item[$\Diamond$] If $2 \le k \le \beta$, we perform $k$ $(j+i,
\uparrow)$-transformations with $i=k-1,... , 0$, in this order.
We get the matrix $\mathbf{M_{3}}$ (fig. 3), and we have for $\ell \in \{0, ... , d-j-k\}$ : $$ u_{j}^k \cdot \det (\Sylv_{j+k, \ell}) = d_{j+k,\ell}. $$

\begin{figure}[htb] \begin{center} \includegraphics{BSP.2}
\end{center} \caption{Shape of the matrix $M_{3}$}
     \end{figure}

\item[$\Diamond$]  If $\beta < k \le \alpha+\beta$, we use $\beta$ $(j+i,
\uparrow)$-transformations with $i=k-1, ..., k-\beta$ in
this order, and $k-\beta$ $(i, \downarrow)$-transformations
with $i=0, ... , k-\beta -1$, in this order. We get the
matrix $\mathbf{M_{4}}$ (fig. 4), and :
                     
\begin{figure}[htb]
\begin{center} \includegraphics{BSP.4} \end{center}
\caption{Shape of the matrix $M_{4}$}
\end{figure}

$$u_{0}^{k-\beta} \cdot u_{j}^\beta \cdot \det
(\Sylv_{j+k, \ell})= d_{j+k,\ell}.$$

\end{itemize}

\end{itemize}

We now prove the theorem, step by step.

\pagebreak

 \textbf{\underline{Proof of (1) :} $2 \le k \le
\alpha+\beta-1$} \bigskip

Since we have to show the nullity of $S_{j+k}$ for $k=2,...  , \alpha
+ \beta -1 $, we need to show that the coefficients $\det(Sylv_{j+k,
\ell})$ vanish for $\ell= 0,...  ,d-j-k$.  This is equivalent to
showing that $d_{j+k, \ell}=0$, for one of the matrices
$\mathbf{M_{1}}$, $\mathbf{M_{2}}$, $\mathbf{M_{3}}$ or
$\mathbf{M_{4}}$, because $u_{0}$ and $u_{j}$ are both different from
zero.

\begin{itemize}

\item   Case $\alpha > 0, \beta > 1$
             
Suppose that $1< \beta \le \alpha$ and $1<k \le \alpha$.  We use
$\mathbf{M_{1}}$ : the submatrix corresponding to $d_{j+k,\ell}$ has
at most one nonzero element on its first row.  We use the
corresponding column to expand it.  The first row of the remaining
minor has only zeros since $\beta \ge 2$.  Hence $d_{j+k, \ell}=0$.
              
If $1<\beta \le \alpha < k \le \alpha+ \beta -1$, we
use $\mathbf{M_{2}}$. 
We have $\alpha+\beta-k \ge 1$ and then 

$$[(2\alpha-k+1)+\beta]-\alpha \ge 2.$$

It follows that there are at least two among the first $\alpha$ rows
 for which at most one entry is nonzero, namely on the $(j+k)$-th column.
  Developing $d_{j+k,l}$ along those two rows shows that it is zero.

              
If $1 \le \alpha < \beta$ and $1< k \le \beta$, we use
$\mathbf{M_{3}}$ to expand $d_{j+k, \ell}$ along the $(j+k)$-th
row, which has at most  one nonzero coefficient. As $\min (k,
\alpha+1) \ge 2$, the row immediately above also has this property, and
we get $d_{j+k, \ell}=0$.
              
Finally, if $1 \le \alpha < \beta < k \le \alpha+\beta-1$, we use
$\mathbf{M_{4}}$. Once again, in $d_{j+k,\ell}$ we have two
successive rows with only one non-zero coefficient,
 on the $(j+k)$-th column (because $(\alpha+1)+(2\beta-k)
\ge \beta+2$).

              
In every case, we see that, if $\alpha >0$ and $\beta >1$, then
$S_{j+k}\equiv 0$. This establishes the first part of 1.

\item Case $\alpha=0, \beta >1$
            
As $2 \le k \le \alpha+\beta-1$, we have $1<k\le \beta$, $\min(k,
\alpha+1)=1$ and we can expand the minor $d_{j+k, \ell}$, using the
rows $j+k,...  , j+1$ in $\mathbf{M_{3}}$, in this order, and then,
using the last $k$ columns.  We obtain, for every $\ell= 0,...,
d-j-k$~: $$ d_{j+k, \ell} = \co_{\ell}(S_{j+1})\cdot
\tc(S_{j+1})^{k-1}\cdot b_{d}^k \cdot S_{j}(0). $$ 

(The factors are written  from left to right, in their order of appearance 
in the successive expansions.) As $d_{j+k, \ell}= u_{j}^k
\det(Sylv_{j+k, \ell})$ and $u_{j}=b_{d}S_{j}(0)$, we have : $$
S_{j}(0)^{k-1} \cdot S_{j+k}= \tc(S_{j+1})^{k-1} \cdot S_{j+1}. $$
            
If $j=0$, $u_{j}=b_{d}$ and $S_{j}(0)$ does not appear in $d_{j+k,
\ell}$. Hence : $$ S_{k}=\tc(S_{1})^{k-1} \cdot S_{1}. $$
            
%
             
\item $\alpha >0, \beta=1$
           
We have $2\le k \le \alpha$ and we use $\mathbf{M_{1}}$, 
expanded along the first $k$ columns, and then along the first $k$
rows. We obtain (the factors appear in  order of  expansions from
the right-hand side of the formula) : 
\begin{eqnarray*} d_{j+k,\ell} &=
&(-1)^{k(j+k+2)}\cdot b_{0}^k\cdot (-1)^{k(j+k+1)}
\co_{j+k-1+\ell}(X^{j}S_{j+1})\\
& & \hspace{7cm} \cdot \lc{(S_{j+1})}^{k-1}
\cdot \lc{(S_{j})}\\
 &=& (-1)^{k}\cdot b_{0}^k  \cdot
\co_{k-1+\ell}(S_{j+1})\cdot \lc{(S_{j+1})}^{k-1}\cdot
\lc{(S_{j})}.  \end{eqnarray*}
          
The result follows.  If $j=0$, the computation is the same : however, in
this case, all the rows of block $B$ collapse.
          
\end{itemize}

\pagebreak
\textbf{\underline{Proof of (2) : $k=\alpha + \beta$}} \bigskip

If $(\alpha,\beta)=(0,1)$, the result is trivial. So, we suppose that
$(\alpha,\beta)\not=(0,1)$. For $j\not=0$, we distinguish two cases.

\begin{itemize}

\item $\beta \le \alpha$
           
We use the  matrix $\mathbf{M_{2}}$ and expand it along the row of
order $\beta$ to obtain : \begin{eqnarray*} d_{j+k, \ell} &= &
u_{0}^{\alpha}\cdot u_{j}^\beta \cdot\det(Sylv_{j+k, \ell})\\
&=& u_{0}^{\alpha} \cdot u_{j}^\beta \cdot (-1)^{n_{0}} \cdot
\co_{j+k-1+\ell}(X^{j+\beta-1} S_{j+1}) \cdot \Delta \\ &=&
u_{0}^{\alpha}\cdot u_{j}^\beta \cdot(-1)^{n_{0}}
\cdot\co_{\alpha+\ell}( S_{j+1}) \cdot\Delta, \end{eqnarray*}
           
where $n_{0}=j+\alpha$ and $\Delta$ is a minor independent of $\ell$.
           
Then, we expand $\Delta$ along the first $\beta-1$  rows, and see
that : $$ \Delta=(-1)^{n_{1}}\cdot \tc(S_{j+1})^{\beta-1}\cdot
\Delta_{1}, $$ with $n_{1}=(j+\alpha)(\beta-1)$. We continue
expanding $\Delta_{1}$ along the first  $\alpha-\beta$  rows ; we have
: $$
\Delta_{1}=(-1)^{n_{2}}\cdot\lc(S_{j+1})^{\alpha-\beta}\cdot\Delta_{2},
$$ with $n_{2}=(j+\alpha)(\alpha-\beta)$. We can then use 
rows $j+1,... , j+\beta$ to compute $\Delta_{2}$ : 
$$
\Delta_{2}=(-1)^{n_{3}} \cdot\lc(S_{j+1})^\beta \cdot\Delta_{3}$$
($n_{3}=\alpha \beta$). Finally, $\Delta_{3}$ can be expanded using
the first $\alpha$ columns and the last $\beta$ ones : $$
\Delta_{3}=(-1)^{n_{4}}\cdot b_{0}^\alpha \cdot b_{d}^\beta \cdot
S_{j}(0), $$ with $n_{4}=j\alpha$. In summary, we have obtained :
$$ u_{0}^\alpha \cdot u_{j}^\beta \cdot \det(Sylv_{j+k, \ell})=
(-1)^N \cdot b_{0}^\alpha \cdot b_{d}^\beta \cdot
\tc(S_{j+1})^{\beta-1} \cdot\lc(S_{j+1})^\alpha \cdot S_{j}(0) \cdot
\co_{\alpha+\ell}(S_{j+1}), $$ with
$N=n_{0}+n_{1}+n_{2}+n_{3}+n_{4}\equiv \alpha(\alpha+\beta) \bmod 2$.
As this computation is valid for every $\ell=0,..., d-j-k$, we have :
$$ \lc(S_{j})^\alpha \cdot S_{j}(0)^{\beta-1} \cdot
S_{j+\alpha+\beta}= (-1)^{\alpha(\alpha+\beta)} \cdot
\lc({S_{j+1})}^\alpha \cdot \tc(S_{j+1})^{\beta-1} \cdot
\frac{S_{j+1}}{X^\alpha}.  $$

\item $\alpha < \beta$
          
We use the same method as in the previous situation, starting with
$\mathbf{M_{4}}$. We expand it along the row of order $j+\beta$ and 
obtain :
          
$$ d_{j+k, \ell}=u_{0}^{\alpha} \cdot u_{j}^\beta \cdot
\det(Sylv_{j+k, \ell})= u_{0}^{\alpha} \cdot u_{j}^{\beta} \cdot
(-1)^{n'_0} \cdot \co_{\alpha+\ell}( S_{j+1}) \cdot \Delta'.$$ 

We expand $\Delta'$ along its rows $j+\beta+1,..., j+k$ to obtain :

$$
\Delta'=(-1)^{n'_{1}} \cdot \lc(S_{j+1})^\alpha \cdot \Delta_{1}',$$

then again along its rows $j+\beta-1,... , j+\alpha+1$ to obtain : 

$$
\Delta_{1}'= (-1)^{n'_{2}} \cdot\tc(S_{j+1})^{\beta-\alpha-1} \cdot
\Delta_{2}',$$
           
and then along its first  $\alpha$ rows : 

$$ \Delta_{2}'=
(-1)^{n'_{3}} \cdot\tc(S_{j+1})^{\alpha} \cdot \Delta_{3}',  $$ 

to finally find that $\Delta_{3}=\Delta_{3}'$.  We now have :
\begin{eqnarray*} n'_{0} &=& \alpha, \\ n'_{1} &=& \alpha^2, \\
n'_{2} &=& \alpha (\beta - \alpha -1), \\ n'_{3} &=& \alpha
(j+ \alpha), \\ n'_{4} &=& n_{4} \; = \; j\alpha .
\end{eqnarray*}
           
We obtain exactly the same final relation as in the case $\beta \le
\alpha$.
           
\end{itemize}

If $j=0$, we have $u_{0}=u_{j}=b_{d}$; and $S_{j}(0)$ disappears at the end of
the successive expansions of the minors. Therefore we get : $$ b_{d}^\alpha
\cdot S_{\alpha+\beta}= (-1)^{\alpha(\alpha+\beta)} \cdot b_{0}^\alpha \cdot\lc
(S_{1})^\alpha \cdot \tc(S_{1})^{\beta-1} \cdot \frac{S_{1}}{X^\alpha}. $$

\medskip \textbf{\underline{Proof of (3) :}} \medskip

Here we cannot use the same transformations of $Sylv_{j+\alpha+\beta+1}$ as above.

We suppose first that $j>0$. Let $R=-\srem (S_{j}, S_{j+1})$ and
$Q=\squo(S_{j},S_{j+1})$. There exist four polynomials $U_{j-1}$,
$V_{j-1}$, $U_{j}$ and $V_{j}$ such that : \begin{eqnarray*} X^{j-1}S_{j} &=&
U_{j-1}A +V_{j-1}B, \\ X^{j} S_{j+1} &=& U_{j}A+V_{j}B \end{eqnarray*}

with $\deg (U_{j-1})\le j-1$, $\deg (V_{j-1})\le j-1$,
$\deg(U_{j})=\deg(V_{j})=j$. We also have : $$ X^\beta R = Q
\frac{S_{j+1}}{X^\alpha} -S_{j}.,$$ 

and deduce that : \begin{eqnarray*}
X^{j+\alpha+\beta}R &=& (QU_{j}-X^{\alpha+1}U_{j-1})A +
(QV_{j}-X^{\alpha+1}V_{j-1})B \\ &=& UA+VB.  \end{eqnarray*}

As $\deg(QU_{j})=j+\alpha+\beta$ and  $\deg(X^{\alpha+1}U_{j-1})\le j+\alpha$,
we have $\deg U = j+ \alpha+ \beta$. Likewise, $\deg V=j+\alpha+\beta$.

Also : $$ \lc(U)=\lc(Q)\lc(U_{j})=\frac{\lc(S_{j})\cdot b_{d} \cdot
S_{j}(0)}{\lc(S_{j+1})}. $$

The equation $X^{j+\alpha+\beta}R=UA+VB$, with $\deg(U)=\deg(V)=j+\alpha+\beta$,
shows that $X^{j+\alpha+\beta}R$ can be obtained as a linear combination of rows
of $Sylv_{j+\alpha+\beta+1}$. As in the previous steps, we transform the row
$j+\alpha+\beta+1$ and obtain a matrix which has the following structure :
\begin{center} \begin{picture}(300,120)(-100,40) \put(-70,160){\line(1,0){215}}
{\thicklines

\put(-69,159){\line(1,0){160}} \put(-69,159){\line(1,-1){50}}
\put(91,159){\line(1,-1){50}} \put(-19,109){\line(1,0){160}}
\put(-16,106){\line(1,0){104}}

\put(-69,104){\line(1,0){160}} \put(-69,104){\line(1,-1){50}}
\put(91,104){\line(1,-1){50}} \put(-19,54){\line(1,0){160}}

\put(-16,51){\line(1,0){160}} } \put(-70,160){\line(0,-1){112}}

\put(-70,48){\line(1,0){215}} \put(145,160){\line(0,-1){112}}

\multiput(-19,153)(0,-7){16}{\line(0,-0){5}}
\multiput(91,153)(0,-7){16}{\line(0,-0){5}}
\multiput(147,106)(7,0){2}{\line(1,0){5}}
\multiput(147,51)(7,0){2}{\line(1,0){5}}

\put(160,140){$_{j+\alpha+\beta}$} \put(160,106){$_{\gets X^{j+\alpha+\beta}
R}$}

\put(160,80){$_{j+\alpha+\beta}$} \put(160,51){$_{\gets X^{j+\alpha+\beta}B}$}

\end{picture} \end{center} \noindent Therefore, for $\ell=0, 1,... ,
d-(j+\alpha+\beta+1)$, we obtain :
$$ \lc (U)\cdot \det(\Sylv_{j+\alpha+\beta+1,\ell})= \underbrace{\left| \matrix{
a_0 & \cdots & a_{j+\alpha+\beta-1} \cr &\ddots &\vdots \cr & & a_0  \cr & &0
\cr b_0&\cdots&b_{j+\alpha+\beta-1}\cr & \ddots & \vdots \cr & & b_0 \cr & & 0 }
\right. }_{j+\alpha+\beta} \matrix { a_{j+\alpha+\beta+\ell} \cr \vdots \cr
a_{\ell-1} \cr \co_\ell(R) \cr b_{j+\alpha+\beta+\ell} \cr \vdots \cr \vdots \cr
b_{\ell}} \underbrace { \left. \matrix {a_d & & & \cr \vdots&\ddots & &\cr
a_{d-j-\alpha-\beta+1} &\cdots&a_d& \cr 0 &\cdots &0 &0\cr b_d & & & \cr \vdots
& \ddots & & \cr \vdots & & \ddots & \cr b_{d-j-\alpha-\beta} & \cdots&\cdots &
b_d  } \right|}_{j+\alpha+\beta+1}.  $$
Expanding these determinants along the
last column, and then along  row $(j+\alpha+\beta+1)$, we see that : 
$$ \lc(U)
\cdot S_{j+\alpha+\beta+1}=b_{d} \cdot S_{j+\alpha+\beta}(0)R. $$ 

We use the value of $\lc(U)$ already computed to obtain the desired result : 

$$ \lc(S_{j})
\cdot S_{j}(0) \cdot S_{j+\alpha+\beta+1}= \lc(S_{j+1})\cdot
S_{j+\alpha+\beta}(0)R. $$

When $j=0$, the polynomials  $U_{j-1}$, $V_{j-1}$, $U_{j}$ and $V_{j}$
are very simple, as we have : $$ S_{0} = 1 . B, \;\; S_{1} = b_{d}A-a_{d}B.$$

The expression of $\lc(U)$ is now : $\lc(U)= \frac{\lc(S_{0})\cdot
b_{d}}{\lc(S_{1})}$. However, the rest of the computation is unchanged, and we
obtain : $$ \lc(S_{0})\cdot S_{\alpha + \beta +1}= \lc(S_{1}) \cdot
S_{\alpha+\beta}(0) R. $$

\end{proof}

\textbf{Remark :} If we define the \textsc{Toeplitz-Bezoutian} of two
monic polynomials $P$ and $Q$ of the same degree as the matrix $Bez(P,Q)$
whose entries are the coefficients of the polynomial 

$$\displaystyle
\frac{P(X)Q^{*}(Y)-P^{*}(Y)Q(X)}{1-XY}.$$

If $sc_{i}(M)$ denotes the $i$-th \textsc{Schur}-complement of the
square matrix $M$ whenever it exists, one can see that we have :

$$S_{i}(0)\lc(S_{i})sc_{i}(Bez(S_{-1},S_{0}))= Bez(S_{i}, S_{i+1}).$$

(See Bini and Pan \cite{BP} p. 169 for the classical result over the Euclidean  remainder sequence. Proof uses same methods). 

\section{Computation of the Symmetric Subresultants Sequence}

The previous theorem gives us a direct method to compute the sequence of
symmetric subresultants of  two polynomials $A$ and $B$, of same degree
$d$ and same valuation 0. It uses symmetric divisions instead of the
determinantal definition.  With  parts 2 and 3 of Theorem \ref{grostheo}, we can 
compute the subsequence $(S_{k_i})_{i=0,...,s}$ ($s\le d$) of the sequence of
the symmetric subresultants,  such that, for each index $i$, the pair
$(S_{k_i},S_{k_{i}+1})$ is $(\alpha_i,\beta_i)$-defective. This implies that,
for each $i$, $S_{k_i}$ is of valuation 0 and degree $d-k_i$ (we have $k_0=0$ as
$S_0=B$). 
Denote by $Q_i$ the $i$-th symmetric quotient of $(S_{k_{i}},
S_{k_{i}+1})$. The sequence $(S_{k_i})_{i=0,...,s}$ is obtained by the
following Euclidean-like algorithm :

\begin{eqnarray*} \lc(S_1)\cdot S_{k_1}(0) \cdot S_{0} &=& Q_0 S_1-\lc(S_0)
\cdot S_{k_1+1}, \\ \lc(S_{k_1+1}) \cdot S_{k_2}(0) \cdot S_{k_1} &=&
Q_1 \frac{S_{k_1+1}}{X^{\alpha_1}} - X^{\beta_1} \lc(S_{k_1})\cdot
S_{k_1}(0)  \cdot S_{k_2+1} ,\\ & \vdots &  \\ \lc(S_{k_s +1}) \cdot
S_{k_s+1}(0)  \cdot S_{k_s} &=& Q_s \frac{S_{k_s+1}}{X^{\alpha_s}}.
\end{eqnarray*}

For such an algorithm, a classical analysis of cost gives a bound of
${\mathcal{O}}(d^2)$ arithmetical operations.  In the important case
of $\Z$, we use \textsc{Hadamard}'s bound for a determinant : if the
size of all the coefficients of the polynomials is bounded by
$\sigma$, then the size of the coefficients of all the $S_{k_{i}}$ is
bounded by $2d(\sigma+\log(d))$.  Therefore, in the case of $\Z$, the
binary cost of the algorithm is in ${\mathcal{O}}(d^2
{\mathcal{M}}(2d(\sigma+\log d)))$ where ${\mathcal{M}}(t)$ denotes
the cost of the multiplication of two integers of absolute value less
than $2^t$.


However, this algorithm can be improved.  In a previous article (see
\cite{BSP}), we studied the case where $B$ is the reciprocal
polynomial of $A$.  In fact the improvement we gave can be applied to
every pair of polynomials $A$ and $B$ in $\D[X]$ of same degree $d$
and valuation zero.  The next section is devoted to showing this.

The ideas we develop here are adaptations to the case of symmetric
subresultants, of ideas already known for ordinary subresultants (see
\cite{LAU}, \cite{L_R},\cite{LR}, \cite{Re}).

\subsection{Transition Matrices}

One idea is to express the transition from a pair \linebreak $(S_{k_{i}},
S_{k_{i}+1})$ to a pair $(S_{k_{i+1}}, S_{k_{i+1}+1})$ with an appropriate
matrix. 

Let $A$ and $B$ be two polynomials in $\D[X]$ of same degree $d$ and valuation
0. Suppose the pair $(S_j, S_{j+1})$ to be $(\alpha, \beta)$-defective ;
set $k=j+\alpha+\beta$ and denote by $Q$ the symmetric quotient of
$\lc(S_{j+1})S_{k}(0)S_j$ by $S_{j+1}$.  With formulae 2 and 3 of the
Structure-Theorem Th. \ref{bigtheo}, we can write, for $j>0$ :

$$ \left( \begin{array}{c } X^{k-1}  S _k \\ X^k S_{k+1} \end{array}  \right)=
M_{j,k} \cdot \left( \begin{array}{ c} X^{j-1}  S _{j} \\ X^j S_{j+1}
\end{array}  \right)$$
           
with \begin{equation} \label{transitmat1}M_{j,k}= \left( \begin{array}{ cc} 0  &
(-1)^{(\alpha+\beta)\alpha} \frac{\lc(S_{j+1})^\alpha
\tc(S_{j+1})^{\beta-1}} {\lc(S_j)^\alpha S_j(0)^{\beta-1}} X^{\beta-1}
\\ -\frac{\lc(S_{j+1})S_k(0)}{\lc(S_j)S_j(0)} X^{\alpha+1} &
\frac{Q}{\lc(S_j)S_j(0)}  \end{array} \right).  \end{equation}

In the case $j=0$, we obtain :

$$ \left( \begin{array}{c } X^{k-1}  S _k \\ X^k S_{k+1} \end{array}  \right)=
M_{0,k} \cdot \left( \begin{array}{ c} S _{0} \\ S_{1} \end{array}
\right)$$ with

\begin{equation} \label{transitmat2} M_{0,k}= \left( \begin{array}{ cc} 0  &
(-1)^{k\alpha} \frac{b_{0}^\alpha \lc(S_{1})^\alpha
\tc(S_{1})^{\beta-1}} {b_d^\alpha} X^{\beta-1}   \\
-\frac{\lc(S_{1})S_k(0)}{b_d} X^{\alpha} & \frac{Q}{b_d}  \end{array}
\right).  \end{equation}

Furthermore, we have (for $j=-1$) : 
$$ \left(  \begin{array}{c } S_0    \\ S_1
\end{array} \right) = \left( \begin{array}{cc } 0 &  1  \\ b_d & -a_d
\end{array} \right) \cdot \left(  \begin{array}{c } A   \\ B   \end{array}
\right).$$ 
                
We can now state a general definition.                 
                
\begin{definition} Let $A=\sum_{i=0}^d a_iX^{i}$ and $B=\sum_{i=0}^d b_iX^{i}$ 
be two polynomials of $\D[X]$ of same degree $d$ and same valuation 0.
Let $(S_i)_{-1 \le i \le d}$ be the sequence of the symmetric
sub-resultants of $A$ and $B$. We denote by $(k_i)_{i=0,...,s}$ (with $k_0=0
< k_1<... <k_s$) the sequence of indices such that $(S_{k_i},
S_{k_i+1})$ is $(\alpha_i, \beta_i)$-defective. 

Then, for $i,j \in \{0,...,s\}$, with $i<j$, we denote by
$M_{k_i,k_j}$ the matrix defined by : $$ M_{k_i,k_j}=
M_{k_{j-1},k_{j}}\cdot M_{k_{j-2},k_{j-1}}\cdot ...  \cdot M_{k_{i},
k_{i+1}},$$

where the matrices $M_{k_\ell,k_{\ell+1}}$ are defined by the above  
formulae
(\ref{transitmat1}) and (\ref{transitmat2}). If $i > 0$, we have :

$$ \left( \begin{array}{c } X^{k_j-1}  S_{k_j} \\ X^{k_j} S_{k_j+1} \end{array}
\right)= M_{k_i,k_j} \cdot \left( \begin{array}{ c} X^{k_i-1}  S _{k_i}
\\ X^{k_i} S_{k_i+1} \end{array}  \right),$$
           
and if $i=0$ : 

$$ \left( \begin{array}{c } X^{k_j-1}  S_{k_j} \\ X^{k_j} S_{k_j+1} \end{array}
\right)= M_{0, k_j} \cdot \left( \begin{array}{ c} S _{0} \\ S_{1}
\end{array}  \right).$$

We call the matrix  $M_{k_i ,k_j}$ the \emph{transition matrix} from the
pair  $(S_{k_i}, S_{k_i+1})$ to the pair $(S_{k_j}, S_{k_j+1})$.
We denote by $M_{k_i}$ the transition matrix from $(A, B)$ to $(S_{k_i},
S_{k_i+1})$ : $$ M_{k_i}=M_{0, k_i} \cdot \left( \begin{array}{cc } 0
&  1  \\ b_d & -a_d  \end{array} \right).$$
          
with the convention that $M_{0, 0}$ is the identity.

\end{definition}

We can now justify the assertion of the previous section : all the
quotients (and remainders) involved in the Structure-Theorem are fraction-free.

\begin{proposition}\label{prop:transitmatrices} Let $A=\sum_{i=0}^d a_iX^{i}$
and $B=\sum_{i=0}^d b_iX^{i}$ be two polynomials of $\D[X]$ of same
degree $d$, and same valuation 0. Let $(S_i)_{-1 \le j \le d}$ be the
sequence of the symmetric sub-resultants of $A$ and $B$. Let $j\in \{1,
..., d-1\}$ be such that $(S_j,S_{j+1})$ is $(\alpha, \beta)$-defective.
Put $k=j+\alpha+\beta$. 


Then the symmetric quotient of $\lc(S_{j+1})S_k(0) S_j$ by $S_{j+1}$
belongs to $\D[X]$, as does the symmetric remainder.


\end{proposition}

\begin{proof}{Proof : }

By Lemma \ref{relbezout} we have for $i>0$ : \begin{eqnarray*} X^{i-1}S_i &=&
U_{i-1}A+V_{i-1}B,\\ X^iS_{i+1} &=& U_i A + V_i B.  \end{eqnarray*}

Therefore, we obtain, for each $j>0$, the following expression of $M_{j}$ :
$$ M_{j}= \left( \begin{array}{cc } U_{j-1} &  V_{j-1}  \\ U_j & V_j
\end{array} \right).  $$

We can directly deduce from  (\ref{transitmat1}) and (\ref{transitmat2}) the value
of $\det(M_{j,k})$. Moreover, if we consider the first line of 
$$ \left(
\begin{array}{c } X^{k-1}  S _k \\ X^k S_{k+1} \end{array}  \right)= M_{j,k}
\cdot \left( \begin{array}{ c} X^{j-1}  S _{j} \\ X^j S_{j+1}
\end{array}  \right),$$ we see that  $\lc(S_k)=
(-1)^{(\alpha+\beta)\alpha}\frac{\lc(S_{j+1})^{\alpha+1}
\tc(S_{j+1})^{\beta-1}}{\lc(S_j)^\alpha S_j(0)^{\beta-1}}.$ Therefore,
we obtain, for $j>0$ : $$
\det(M_{j,k})=\frac{\lc(S_k)S_k(0)}{\lc(S_j)S_j(0)}X^{\alpha+\beta},$$
and  $j=0$ yields : $$
\det(M_{0,k})=\frac{\lc(S_k)S_k(0)}{b_d}X^{\alpha+\beta-1}.$$ 

As above, we denote by $(k_i)_{0 \le i \le m}$  the indices
such that $(S_{k_i}, S_{k_i+1})$ is $(\alpha_i, \beta_i)$-defective with
$k_0=0$ and $k_m=j$. We have : 

\begin{eqnarray*} \det(M_j) &=& \det(M_0)\cdot\left(  \prod_{i=0}^{m-1} \det (
M_{k_i,k_{i+1}}) \right), \\ &=&- b_d \cdot \prod_{i=0}^{m-1}
\det(M_{k_i,k_{i+1}}),  \\ &=& -b_d \cdot
\frac{\lc(S_{k_1})S_{k_1}(0)}{b_d}X^{\alpha_0+\beta_0-1} \cdot
\prod_{i=1}^{m-1}
\frac{\lc(S_{k_{i+1}})S_{k_{i+1}}(0)}{\lc({S_{k_i})S_{k_i}(0)}}X^{\alpha_i+\beta_i},
\\ &=&- \lc(S_{k_m})S_{k_m}(0)X^{k_1-k_0-1}\cdot \prod_{i=1}^{m-1}
X^{k_{i+1}-k_i}, \\ &=& - \lc(S_{k_{j}}) S_{k_{j}}(0) X^{j-1}.
\end{eqnarray*}

Consequently, the matrix $M_{j}$ is invertible and we easily see that, if $j>0$ : $$
\det(M_j)M_j^{-1}=\left( \begin{array}{cc } V_j &  -V_{j-1}  \\ -U_j & U_{j-1}
\end{array} \right).$$

When $j=0$, we get :   $b_dM_0^{-1}=\left( \begin{array}{cc } a_d &  1
\\ b_d & 0  \end{array} \right).$

By definition of $M_j$, we have for $0 \le j< k$, $M_{j,k}=M_k \cdot M_j^{-1}$.
Then  for $j>0 :$ $$ -\lc(S_j) S_j(0) X^{j-1} M_{j,k}= \left(
\begin{array}{cc } U_{k-1} &  V_{k-1}  \\ U_k & V_k  \end{array} \right) \cdot
\left( \begin{array}{cc } V_j &  -V_{j-1}  \\ -U_j & U_{j-1}
\end{array} \right), $$ 

and for 
$j=0$ : $$ - b_d M_{0,k}= \left( \begin{array}{cc } U_{k-1} &
V_{k-1}  \\ U_k & V_k  \end{array} \right) \cdot \left( \begin{array}{cc
} a_d & 1  \\ b_d & 0  \end{array} \right). $$

Identifying the bottom right-hand side entries of these matrices,
yields \linebreak if $j>0 :$
$$ X^{j-1} Q =U_{j-1}V_k-U_kV_{j-1} \in \D[X],$$ 
and $Q=-U_{k}$ when $j=0$.  \end{proof}

\subsection{Symmetric truncation}

The computation of the symmetric quotient of two polynomials does not
involve all of their coefficients.  In fact, we only need the leading
and trailing terms of the divisor.  More generally, the computation of
successive symmetric quotients can be done with only the knowledge of
a few leading and trailing terms of the first divisors.  This way it
appears cheaper to compute successive quotients instead of successive
remainders, as we use only small parts, which we will refer to as
``symmetric truncation" of the polynomials.


First we define the symmetric truncation of a polynomial.

\begin{definition} Let $P=\sum_{i=0}^d p_iX^{i} $ be an element of 
    $\D[X]$, $P\neq 0$. For
$\ell \in \{1,... , \lfloor d/2 \rfloor \}$, we denote by $P_{|\ell}$ the
polynomial 
$$ P_{|\ell}=p_0+\cdots +p_{\ell-1}X^{\ell-1}+
p_{d-\ell+1}X^{\ell}+ \cdots + p_d X^{2\ell-1}.$$ 
For $\ell=0$, we write $P_{|0}=0$, and for $\ell > \lfloor d/2 \rfloor $, $P_{|\ell}=P$.
\end{definition}

We now analyse the cases where truncation of two polynomials does not affect
their symmetric quotient.

\begin{lemma}\label{lem1} Let $P$ and $P_1$ be two polynomials of $\D[X]$ such
that  $\deg (P)=d$, $\deg(P_1)=d-\beta\le d$, $v(P)=0$ and
$v(P_1)=\alpha \ge 0$. Then, 
 $$ \squo(P, P_1) = \squo
(P_{|(\alpha+\beta+1)}, {P_1}_{|(\alpha+\beta+1)}), $$ where $P_1$ is
considered as a polynomial of degree $d$  in order to compute its truncation.
\end{lemma}

\begin{proof}{Proof :} Set $\widehat P = P_{|(\alpha+\beta+1)}$,
$\widehat P_1={P_1}_{|(\alpha+\beta+1)}$ and $\gamma=
d-2(\alpha+\beta)-1$. We have $\deg \widehat P=2(\alpha+\beta)+1$, $\deg
\widehat P_1 = 2 \alpha+\beta +1$, $v(\widehat P)=0$, and $v(\widehat
P_1)=\alpha$. Then, let us consider the following symmetric divisions
:
\begin{eqnarray*} P &=& Q \frac{P_1}{X^\alpha} + X^\beta R \;\;\; \mathrm
{with}\;\;\; \deg(R)<d-\alpha-\beta,  \\ \widehat P &=& \widehat Q
\frac{\widehat P_1}{X^\alpha} + X^\beta \widehat R \;\;\; \mathrm
{with}\;\;\; \deg(\widehat R)<\alpha+\beta+1.  \end{eqnarray*}

We have $\deg(Q)=\deg(\widehat Q)=\alpha+\beta$ and we can write :
$Q=Q_1X^\beta+Q_2$ and $\widehat Q=\widehat Q_1X^\beta+ \widehat Q_2$, where
$\deg Q_{2}$ and $\deg \widehat Q_{2}$ are strictly less than $\beta$, and
$\deg Q_{1}=\deg \widehat Q_{1}=\alpha$.

Then :

$$ P-X^\gamma \widehat P= Q \frac{P_1-X^\gamma \widehat P_1}{X^\alpha} +
(Q-\widehat Q) \frac{\widehat P_1 X^\gamma}{X^\alpha}+ X^\beta(R-X^\gamma
\widehat R).$$

Since $\deg (P-X^\gamma \widehat P)$  and $\deg (P_{1}-X^\gamma \widehat
P_{1})$ are less than $d-\alpha-\beta$, we see that $\deg(Q- \widehat Q) <
\beta$, and therefore, $Q_1= \widehat Q_1$. 
Similarly, we compare the valuation of both sides at the identity : 
$$ P- \widehat P= Q \frac{P_1-\widehat P_1}{X^\alpha} + (Q- \widehat Q)
\frac{\widehat P_1}{X^\alpha} + X^\beta (R-\widehat R).$$ As $v(P-\widehat P) >
\alpha+\beta$ and $v((P_1-\widehat P_1)/X^\alpha) > \beta$, we conclude that
$v(Q-\widehat Q) \ge \beta$, \textit{i.e.} $Q_2= \widehat Q_2$. Hence $Q=
\widehat Q$.

\end{proof}

We can also compare the truncation of the symmetric subresultants of two
polynomials with the symmetric subresultants of their truncations.

\begin{lemma}\label{lem2}

Let $A$ and $B$ be in $\D[X]$ of same degree $d$ and valuation 0. Let
$(S_{i})_{-1 \le i\le d}$ be the sequence of the symmetric subresultants of $A$
and $B$. Let, $({\widehat S_j})_{-1 \le j \le 2\ell-1}$ be the sequence of the
symmetric subresultants of $A_{|\ell}$ et  $B_{|\ell}$   ($\ell$ fixed in $\{1,
..., \lfloor d/2 \rfloor \}$). Then for $1 \le j< \ell$, we have :
$$ {S_{j|(\ell-j)}}=  {\widehat S_{j|(\ell-j)}}.$$

\end{lemma}

\begin{proof}{Proof : } The proof is based on the definition of the coefficients of the
symmetric subresultants. Set  $A=\sum_{i=0}^d a_iX^{i}$ and
$\widehat A = \sum_{i=0}^{2\ell-1} \hat a_iX^{i}$ ( respectively
$B=\sum_{i=0}^d b_iX^{i}$ and $\widehat B= \sum_{i=0}^{2\ell-1} \hat
b_iX^{i}$). For $0\le k < \ell-j$, the coefficient of order $k$ of
$\widehat S_{j|\ell - j }$ is given by :

\begin{eqnarray*} \co_{k}(\widehat S_{j| \ell -j}) & = & \left|
\begin{array}{ccccccc}
\hat a_{0} & \ldots & \hat a_{j-2} & \hat a_{k+j-1} &\hat  a_{2 \ell -1}
&  &   \\

& \ddots & \vdots & \vdots  & \vdots & \ddots &   \\

&  & \hat a_{0} & \vdots & \vdots &  &   \\

& & 0 &\hat a_{k} & \hat  a_{2 \ell -j} & \ldots & \hat  a_{2 \ell -1}
\\
\hat b_{0} & \ldots & \hat b_{j-2} & \hat b_{k+j-1} &\hat  b_{2 \ell -1} &  &
\\

& \ddots & \vdots & \vdots  & \vdots & \ddots &   \\

&  & \hat b_{0} & \vdots  & \vdots &  &   \\

& & 0 &\hat b_{k} & \hat  b_{2 \ell -j} & \ldots & \hat  b_{2 \ell -1}
\\

    \end{array} \right| \\ &=& \left| \begin{array}{ccccccc}
a_{0} & \ldots &  a_{j-2} &  a_{k+j-1} &  a_{d} &  &   \\

& \ddots & \vdots & \vdots  & \vdots & \ddots &   \\

&  &  a_{0} &  \vdots & \vdots &  &   \\

& & 0 & a_{k} &  a_{d-j+1} & \ldots &  a_{d}  \\
b_{0} & \ldots &  b_{j-2} &  b_{k+j-1} & b_{d} &  &   \\

& \ddots & \vdots &   \vdots & \vdots & \ddots &   \\

&  & b_{0} &  \vdots & \vdots &  &   \\

& & 0 & b_{k} &  b_{d-j+1} & \ldots &  b_{d}  \\

    \end{array} \right| = \co _k(S_{j | \ell - j}).  \end{eqnarray*}

In the same way, if $ \ell -j  \le k < 2\ell- 2j$, we have :

\begin{eqnarray*} \co_{k}(\widehat S_{j | \ell - j}) & = &\co_{k+j}(\widehat
S_{j}),\\ &=& \left| \begin{array}{ccccccc}
\hat a_{0} & \ldots & \hat a_{j-2} & \hat a_{k+2j-1} &\hat  a_{2 \ell
-1} &  &   \\

& \ddots & \vdots &  \vdots  & \vdots & \ddots &   \\

&  & \hat a_{0} &  \vdots & \vdots &  &   \\

& & 0 &\hat a_{k+j} & \hat  a_{2 \ell -j} & \ldots & \hat  a_{2 \ell
-1}  \\
\hat b_{0} & \ldots & \hat b_{j-2} & \hat b_{k+j-1} &\hat  b_{2 \ell -1} &  &
\\

& \ddots & \vdots &   \vdots & \vdots & \ddots &   \\

&  & \hat b_{0} &  \vdots & \vdots &  &   \\

& & 0 &\hat b_{k} & \hat  b_{2 \ell -j} & \ldots & \hat  b_{2 \ell -1}
\\

    \end{array} \right| \\ &=& \left| \begin{array}{ccccccc}
a_{0} & \ldots &  a_{j-1} &  a_{d-2\ell+k+2j} &  a_{d} &  &   \\

& \ddots & \vdots &   \vdots & \vdots & \ddots &   \\

&  &  a_{0} &  \vdots & \vdots &  &   \\

& & 0 & a_{d-2\ell+k+j+1} &  a_{d-j+1} & \ldots &  a_{d}  \\
b_{0} & \ldots &  b_{j-1} &  b_{d-2\ell+k+2j} & b_{d} &  &   \\

& \ddots & \vdots &   \vdots & \vdots & \ddots &   \\

&  & b_{0} &  \vdots & \vdots &  &   \\

& & 0 & b_{d-2\ell+k+j+1} &  b_{d-j+1} & \ldots &  b_{d}  \\

    \end{array} \right| \\ &=& \co _{d-2\ell+k+j+1}(S_{j}) = \co _{k}(S_{j
    |(\ell-j)}).  \end{eqnarray*}

Therefore ${S_{j|(\ell-j)}}$ and  ${\widehat S_{j|(\ell-j)}}$ have the same
coefficients.  \end{proof}

As a consequence, we have $S_{j|k}=\widehat S_{j|k}$ for every $k$ such that
$0\le k\le l-j$. Also $S_j(0)=\widehat S_j(0)$ and $\lc (S_j) = \lc
( \widehat S_j) $ for every $j< \ell$.

Further, for a given $\ell$, we can predict how many symmetric quotients will be
preserved if we replace  $A$ and $B$ by $A_{| \ell}$ and $B_{| \ell}$ in the
computations.

\begin{theorem} \label{theo:divandconq} Let $A$ and $B$ be in $\D[X]$ of same
degree $d\ge 4$ and valuation 0.  Let $(S_{i})_{-1 \le i\le d}$ be the
sequence of the symmetric subresultants of $A$ and $B$. For $\ell \in
\{2, ..., \lfloor d/2 \rfloor \}$, let $(\widehat S_j)_{-1 \le j \le
2\ell-1}$ be the sequence of the symmetric subresultants of $A_{|\ell}$
et  $B_{|\ell}$.

Let $(k_i)_{0 \le i \le s}$, respectively $(\widehat k_i)_{0 \le i \le s'}$, be the
indices such that the pairs $(S_{k_i}, S_{k_i+1})$, respectively $(\widehat
S_{\widehat k_i}, \widehat S_{\widehat k_i+1})$, are $(\alpha_i,
\beta_i)$-defective, respectively $(\widehat \alpha_i, \widehat
\beta_i)$-defective (we have $k_0=\widehat k_0=0$).

For each $i$ such that $ S_{ k_i+1}\not=0$, set $Q_i=\lc(S_{k_i +1})
S_{ k_{i +1}} (0) \squo(S_{ k_i}, S_{ k_i+1})$ and for each $i$ such
that $\widehat S_{\widehat k_i+1}\not=0$, set  $\widehat Q_i=\lc(\widehat
S_{\widehat k_i +1}) \widehat S_{\widehat k_{i +1}} (0)$ 
\linebreak $
\squo( \widehat S_{\widehat k_i}, \widehat S_{\widehat k_i+1})$.  Then
$M_{ k_i, k_{i+1}}$, respectively $\widehat M_{\widehat k_i, \widehat
k_{i+1}}$, are the transition matrices of the sequence $(S_{j})_{1\le
j \le d}$, respectively $(\widehat S_j)_{1\le j \le 2\ell -1}$.

Let $m$ be an index such that $1 \le m \le s$ and let $k_m +1 < \ell$,
then for all $i=0, 1, ...  , m-1$, we have : $$ \alpha_i= \widehat
\alpha_i,
\;\;\beta_i=\widehat \beta_i, \;\; Q_i = \widehat Q_i , 
k_{i}=\widehat k_{i},$$
and finally, $\widehat M_{\widehat k_i, \widehat k_{i+1}}= M_{ k_i,  k_{ i+1}}$.
\end{theorem}

\begin{proof}{Proof : }

First notice  that for any $i=0,..., m-1$, we have $k_{i+1}=k_i+\alpha_i +
\beta_i$; it follows that : $$ \sum_{i=0}^{m-1} \alpha_i +\beta_i < \ell.$$
For each $j<\ell$, by Lemma \ref{lem2}, we have $S_{j|\ell-j}=\widehat
S_{j|\ell-j}$. Therefore, for each $j=1,2,..., \ell- 1$, we have
$S_{j}(0)=\widehat S_{j}(0)$ as well as $\lc(S_{j})=\lc(\widehat S_{j})$. Then, we see that $k_{i}= \widehat k_{i}$ for
every $i=0, 1,... , m$. Furthermore, as $k_{i+1}- k_{i}=\alpha_{i}+\beta_{i}$,
and $\widehat k_{i+1}-\widehat k_{i}=\widehat \alpha_{i}+ \widehat \beta_{i}$,
we have $\alpha_{i}+\beta_{i}=\widehat \alpha_{i} + \widehat \beta_{i}$ for
every $i=0, 1,... , m-1$.

We claim that $\alpha_{i}=\widehat \alpha_{i}$ ($i=0,\ldots, m-1$).
This will also imply that $\beta_{i}=\widehat \beta_{i}$ for each
$i=0,1,..., m-1$. Indeed, we have $S_{k_{i}+1 | \ell-k_{i}-1} = \widehat
S_{k_{i}+1 | \ell-k_{i}- 1}$.  Therefore, the $\ell-k_{i}-1$ bottom coefficients
of $S_{k_{i}+1}$ and $\widehat S_{k_{i}+1}$ are equal. But
$v(S_{k_{i}+1})=\alpha_{i}$ and we have  $k_{i}+\alpha_{i}+\beta_{i}=k_{i+1} \le
k_{m}< \ell$. Thus $\alpha_{i}$ is less than $\ell -k_{i}-\beta_{i} \le \ell -
k_{i}-1$. The valuations of  $S_{k_{i}+1}$ and $\widehat S_{k_{i}+1}$ must then be
equal.

Having proved that the sequences of indices $(k_{i})_{0 \le
i < m}$, $(\alpha_{i})_{0 \le i < m}$, $(\beta_{i})_{0 \le i < m}$ are equal to their
counterparts, we now show the equality of the symmetric quotients.

First we have, by Lemma \ref{lem1} :
\begin{eqnarray*} Q_{i} & = & \squo \bigl( \lc (S_{k_{i}+1})S_{k_{i+1}}(0)\cdot
S_{k_{i}}, S_{k_{i}+1}  \bigr) \\ & = & \squo \bigl( \lc
(S_{k_{i}+1})S_{k_{i+1}}(0)\cdot S_{k_{i}| \alpha_{i}+\beta_{i}+1},
S_{k_{i}+1 | \alpha_{i}+\beta_{i}+1}  \bigr), \end{eqnarray*} 
since $(S_{k_{i}}, S_{k_{i}+1})$ is $(\alpha_{i}, \beta_{i})$-defective . 

If $i<m$ and $k_{m}+1<\ell$, we have $\alpha_{i}+\beta_{i}+1 < \ell -
k_{i}$, and, by Lemma \ref{lem2}, $S_{k_{i}| \alpha_{i}+\beta_{i}+1}=
\widehat S_{k_{i} | \alpha_{i}+\beta_{i}+1}$.  In respect of
$S_{k_{i}+1 | \alpha_{i}+\beta_{i}+1}$, the truncature is applied to
$S_{k_{i}+1}$ considered of formal degree $d-k_{i}$ (Lemma
\ref{lem1}).  But, by Lemma \ref{lem2}, we have $ S_{k_{i}+1 |
\alpha_{i}+\beta_{i}+1}= \widehat S_{k_{i}+1 |
\alpha_{i}+\beta_{i}+1}$, polynomials being truncated with their actual
degree.  However using formal degree $d-k_{i}$ instead of actual
degree $d-k_{i}-\beta_{i}$, we do not take into account  so many
coefficients and the equality of the truncatures holds as well.

Since the leading coefficients and constant terms of the sequence
$(S_{j})_{0\le j <\ell}$ and $(\widehat S_{j})_{0\le j <\ell}$ are
equal, we can write : \begin{eqnarray*} Q_{i} & = & \squo \bigl( \lc
(\widehat S_{k_{i}+1}) \widehat S_{k_{i+1}}(0)\cdot \widehat S_{k_{i}|
\alpha_{i}+\beta_{i}+1}, \widehat S_{k_{i}+1 | \alpha_{i}+\beta_{i}+1}
\bigr) \\ & = & \widehat Q_{i}.
 \end{eqnarray*}
Finally, inspecting the expression of the transition matrix $M_{k_{i},
k_{i+1}}$ given by (\ref{transitmat1}) and (\ref{transitmat2}), we see that all
the ingredients have been proven to be equal for the two matrices  $M_{k_{i},
k_{i+1}}$ and $\widehat M_{k_{i}, k_{i+1}}$ ($i=0, \ldots, m-1$).

\end{proof}

\subsection{Fast Algorithm}
We now describe the \textbf{FSSR} Algorithm which is written in
pseudo-code further down.

Let $A$ and $B$ be two polynomials of $\D[X]$ of same degree and
valuation 0.  They are considered as global variables.  The
\textbf{FSSR} Algorithm takes as input a pair $(S_{k_{i}},
S_{k_{i}+1})$ of two successive symmetric subresultants of $A$ and
$B$, $(\alpha_i, \beta_i)$-defective and an integer $r< (d-k_{i})$.

It returns the sequence of the symmetric quotients $(Q_{j},
\alpha_{j}, \beta_{j})_{i\le j < v-1}$ with $v$ the largest index such
that $k_{v}  < k_{i}+r $.  It returns also the transition matrix
$M_{k_i, k_{v}}$.

In the general case, we are interested in finding the entire sequence
of symmetric quotients of $A$ and $B$, and \textbf{FSSR}$(S_0, S_1,
d)$ with $S_0=B$, $S_1=\lc(B)A-\lc(A)B$ will suffice.  This way, we
compute the entire sequence of symmetric quotients except perhaps for
the last one which can be obtained with an extra division.

How does this work ?  We use a strategy of \emph{divide and conquer},
to compute a partial sequence at each step.  Here is a description of
each non-trivial step.


Step 1 : If $S_{k_{i}+1}$ is 0, we have already reached the end of the 
sequence of the symmetric subresultants of $A$ and $B$.

Step 2 : If $r\le 2$, the algorithm performs symmetric divisions starting
with the polynomials $S_{k_{i}\mid r}$ and $S_{k_{i}+1 \mid r}$ 
whose degrees are at most 3. It computes also directly the 
corresponding transition matrix.

Step 4 : a call to \textbf{FSSR} $\left( S_{k_i |  r }, S_{k_i +1
|  r }, \lceil \frac{r}{2} \rceil \right)$ is
executed.

Since the third recursive call, the coefficient of truncature is
stricktly lower than $ \lfloor \frac{d-k_i}{2}\rfloor $, and therefore
 Theorem \ref{theo:divandconq}  can be  applied : the algorithm computes
$Q_{j}$, $\alpha_{j}$, $\beta_{j}$ for $j=i, \ldots , u-1$ as well as
$M_{k_i, k_{u}}$, with $u$ the largest index such that $k_{u} < k_i +
\lceil r/2 \rceil $.

Step 5 : We compute $S_{k_{u}}$, and $ S_{k_{u}+1}$ via
$M_{k_i, k_{u}}$.

Step 6 : Then, via a symmetric quotient, we compute $Q_{u}$ and add it
to the list of quotients already computed.  $M_{k_{i},k_{u+1}}$ is
computed as well as $(S_{k_{u+1}},S_{k_{u+1}+1})$.

This intermediary step is needed to guarantee that  the coefficient of truncature in the next call
to \textbf{FSSR} (step 7) is smaller than $\lceil \frac{r}{2} \rceil$.

Step 7 : We perform a second call to \textbf{FSSR}$\left(
S_{k_{u+1}\mid r} , S_{k_{u+1} +1\mid r}, r-(k_{u+1}-k_{i}) \right)$.
We therefore obtain symmetric quotients $Q_{u}$ up to $Q_{v-1}$ with
$v$ the largest index such that $k_v+1<r+k_i$.

Step 8 : We get together the pieces already computed.

\bigskip


\begin{figure}[!btp] {{\scriptsize \begin{tabular}{lp{10cm} } \hline \multicolumn{2}{c}
{ALGORITHM \textbf{FSSR}}\rule[-3pt]{0pt}{18pt} \\ \hline \hline
\textbf{INPUT :} \rule[-3pt]{0pt}{18pt}& 
-- $(S_{k_{i}}, S_{k_{i}+1})$,
a pair $(\alpha_{i}, \beta_{i})$-defective of symmetric \\
 & \hspace{3ex} subresultants of $A$, $B$,\\

&-- $r$  a positive integer, $r \le d-k_i $.\\ 

\textbf{OUTPUT :} &-- the list $L:=[Q_{i}, \alpha_{i}, \beta_{i}, ...,
Q_{v-1}, \alpha_{v-1}, \beta_{v-1}]$ and $M_{k_i, k_{v}}$, where $v$
is the biggest integer such that $k_v < r +k_i$.
\rule[-7pt]{0pt}{12pt}\\
\\
\hline
\\

\textbf{MAIN PART :} 
&1 -- IF $S_{k_{i}+1}=0$  then RETURN $L:=[] $, and $M:=Id_{2}$.  
\\
&2 -- ELSE IF $r \le 2 $ then compute $L$ using symmetric divisions 
 of $S_{k_{i}\mid r}$ with $S_{k_{i}+1 \mid r }$ and 
 $M_{k_{i},k_u}$ from definition.\\
& \quad  -- ELSE \\
&3 -- $r ':= \lceil \frac{r}{2}\rceil $;\\ 
&4 -- $L_{1}:=$\textbf{FSSR}${(S_{k_{i} | r}, S_{k_{i}+1| r}, r ')}$; \\ 
&\hspace{1cm} \% \textit{$L_{1}$ contains :}\\ 
&  \hspace{1cm} \% $Q_{i}, \alpha_{i},
\beta_{i}, ..., Q_{u-1}, \alpha_{u-1}, \beta_{u-1}$,\\
& \hspace{1cm} \% \textit{we get also} : $M_{k_{i}, k_{u}} $,\\ 
& \hspace{1cm} \% \textit{with $u$, largest integer such that} 
$k_{u} <  r '+k_i$.\\

&5 -- Compute $S_{k_{u} } $ and $S_{k_{u}+1} $ by :\\ 
& $$ \left( \begin{array}{c}
X^{k_{u}-1} S_{k_{u}} \\ X^{k_{u}} S_{k_{u}+1} \end{array} \right) =
M_{k_i, k_{u}} \cdot 
\left( \begin{array}{c} X^{k_{i}-1}S_{k_i} \\ X^{k_{i}}S_{k_i+1} \end{array} \right).$$ \\ 

&6 -- 
$Q_{u}=\lc(S_{k_{u}+1})S_{k_{u+1}}(0)\squo(S_{k_{u}},S_{k_{u}+1})$ ; 
\\
& \hspace{0.5cm} $L_{1}=L_{1} \bigcup \{ Q_{u} \}$. 
$M_{k_i, k_{u+1}}=M_{k_u, k_{u+1}}\cdot M_{k_i, k_{u}}$\\
& \quad  Compute $S_{k_{u+1} } $ and $S_{k_{u+1}+1} $ by :\\
& $$ \left( \begin{array}{c}
X^{k_{u+1}-1} S_{k_{u+1}} \\ X^{k_{u+1}} S_{k_{u+1}+1} \end{array} \right) =
M_{k_{i}, k_{u+1}} \cdot 
\left( \begin{array}{c} X^{k_{i}-1}S_{k_i} \\ X^{k_{i}}S_{k_i+1} \end{array} \right).$$ \\ 

&7 -- $L_{2}:=$\textbf{FSSR}$(S_{k_{u+1}\mid r } , S_{k_{u+1}+1\mid r}, 
r-(k_{u+1}-k_{i}))$;\\ 
& \hspace{1cm} \% \textit{$L_{2}$ contains :} \\ 
& \hspace{1cm} \% $Q_{u+1}, \alpha_{u+1}, \beta_{u+1}, ...,
Q_{v-1}, \alpha_{v-1}, \beta_{v-1} ;$ \\
& \hspace{1cm} \% \textit{we get also} : $M_{k_{u+1}, k_{v}} $.\\
& \hspace{1cm} \% with $v$, largest integer such that $k_{v} < r+k_{u+1}$.\\

&8 -- $L:=L_{1} \bigcup L_{2}$ ; $M_{k_i,k_{v}}=M_{ k_{u+1},k_{v}} \cdot M_{k_{i}, k_{u+1}} $ \\ 

\\

\textbf{END.}&\\ \hline

\end{tabular} }} \end{figure}

Remark : throughout the algorithm, instead of computing
$M_{k_{i},k_{m}}=M_{k_{j},k_{m}}\cdot M_{k_{i},k_{j}}$ for $0\le
i<j<m\le s$, it is preferable to compute :

$$M_{k_i,k_{m}}=\bigg(\Big(\big(\lc(S_{k_{j}}) S_{k_{j}}(0)\big)\cdot 
M_{k_j,k_{m}}\Big)\cdot M_{k_i,k_{j}}\bigg)/(\lc(S_{k_{j}}) S_{k_{j}}(0)) $$

using the order of operations indicated by the parentheses. In doing
so, we keep all computations in $\D[X]$ and the algorithm remains
fraction-free.


We now consider its cost.

\begin{theorem}\label{costFSSR} Let $\D$ be a sub-ring of $\C$ and let $A$ and
$B$ be two polynomials of same degree $d$ in $\D[X] $. The
algorithm \textbf{FSSR}$(S_{0}, S_{1},d)$ with $S_{0}=B$ and
$S_{1}=\lc(B)A-\lc(A)B$ uses at most \begin{displaymath}
{\mathcal{O}}({\mathcal{M}}(d).\log(d))={\mathcal{O}}(d
\log^2(d) \log \log(d)) \end{displaymath} arithmetical operations in $\D$
(${\mathcal{M}}(d)$ denotes the cost in arithmetical operations
of multiplying two polynomials of degree at most $d$ in $\D[X]$).

If $A$ and $B$ are elements of $\Z[X]$ or $\Z[i][X]$, and if the size of their
coefficients  is bounded by $\sigma$, then \textbf{FSSR}$(S_{0},
S_{1},d)$ is executed in less than \begin{displaymath}
{\mathcal{O}}\Big((d^2.(\sigma+\log(d)).\log(d\sigma+d\log(d)).\log
\big(
\log(d\sigma+d\log(d))\big).\log(d)\Big) \end{displaymath} binary operations on a
multiband {\sc Turing} machine, using DFT.  \end{theorem}

\begin{proof}{Proof : } Let us denote by ${{\mathcal{CF}}(\delta)}$ the
cost in terms of arithmetical operations of the computation of
$\mathbf{FSSR}(S_{0}, S_{1},\delta)$.
We do not take into account the degrees of the polynomials $S_{0}$ and
$S_{1}$, as, from the very beginning of the algorithm, these
polynomials are truncated to order $\delta$ and the degrees of the
polynomials that we really manipulate are lower than $2 \delta -1$.

During the execution of $\mathbf{FSSR}(S_{0}, S_{1},\delta)$, we use
two calls of $\mathbf{FSSR}$ with $\delta$ replaced by $\lceil
\frac{\delta}{2}\rceil$.  The intermediate computation consists of
some multiplications and a symmetric division : the number of
arithmetical operations is bounded by
${\mathcal{O}}\big({\mathcal{M}}(\delta)\big)$ .  Therefore, we have :

$$ {\mathcal{CF}}(\delta) \le 
2{\mathcal{CF}}\bigg(\bigg\lceil\frac{\delta}{2}\bigg\rceil \bigg) +
{\mathcal{O}}\big({\mathcal{M}}(\delta)\big).$$

It follows that ${\mathcal{CF}}(\delta)$ is bounded by
${\mathcal{O}}\left({\mathcal{M}}(\delta)\log(\delta)\right)$. Hence 
the first assertion with $\delta=d$.  

In the case of $\Z$ or $\Z[i]$, we follow the same arguments.
However, we have to bound the size of the coefficients appearing in
the algorithm.  These coefficients are minors of $\Sylv_{d}(A, B)$.
They can be bounded by \textsc{Hadamard}'s formula : their size is
less than $\tau=2d(\sigma+\log(d))$.  The coefficients of the
transition matrices $M_{k_i, k_{j}}$ are of the same size.  If
${{\mathcal{M}}(d, \tau)}$ is the binary cost to compute the product
of two polynomials of degree less than $d$ with coefficients of size
bounded by $\tau$, we get : $$ {\mathcal{CF}}(d,
\sigma) \le {\mathcal{O}}\left({\mathcal{M}}(d, \tau) \log(d)\right).$$
This proves the result in the case of a multiband \textsc{Turing} machine.
\end{proof}

Remark : it might surprise the reader that we compute the sequence of
symmetric quotients instead of the symmetric sub-resultants.  Indeed
as far as applications are concerned the important elements are the
symmetric remainders and not the symmetric quotients.  In fact, the
applications we know of use either the constant terms of a sequence of
symmetric remainders, or a particular symmetric remainder.  When the
sequence of symmetric quotients is known, the sequence of $S_{k_i}(0)$
can be computed in ${\mathcal{O}}(d)$ as we can see in the
introduction to Part 4.

In this  case, when a particular symmetric remainder is needed, computing
the corresponding transition matrix is enough to
determine this specific remainder, up to a few additional
operations.

%

\section{Application to  \textsc{Toeplitz} matrices}

In this section we consider the relationship between sequences of  principal
minors of a \textsc{Toeplitz} matrix and of the symmetric
sub-resultants of polynomials. As a consequence, we will get new algorithms to
compute the signature and the inverse of such a matrix. We do not improve the
cost of algorithms  presented in \cite{BGY} and \cite{GE1} and already used
in the complex numerical case. However, in the case of integer coefficients, we
control the size of results and use fraction-free computations;  this is well
suited for computer algebra.

\subsection{Relationship between \textsc{Toeplitz} matrices and symmetric
sub-resultants}

We first establish a link between  constant terms of the symmetric
subresultants and principal minors of a \textsc{Toeplitz}  matrix.

\begin{proposition}\label{linkssrtop} Let $F=\sum_{i=0}^{d}f_{i}X^{i}$ and
$G=\sum_{i=0}^{d}g_{i}X^{i}$ be two polynomials of equal valuation; we suppose that the
degree of $F$ is exactly $d$;  the degree of $G$ is formally considered equal to
 $d$ but could be less. Let 
$$ \frac{G}{F}= v+\sum_{i\ge
1}v_{i}X^{i} $$ 
be the expansion around zero of $G / F$, and 
$$
\frac{G}{F}= -u-\sum_{i\ge 1}u_{i}X^{-i} $$ 
 its expansion around
infinity. Let ${\mathcal{T}}_{k}(F,G)=(t_{i,j})_{1\le i,j,\le k}$ be the
\textsc{Toeplitz} matrix  : $$ \left\{ \begin{array}{ccc}

t_{i,j}& = & v_{j-i}\;\;\; \mathrm{if}\; \;\; i<j \\

t_{i,j}& = & u_{i-j} \;\; \; \mathrm{if}\; \;\; i>j \\

t_{i,j}& = & u+v \;\; \; \mathrm{if}\; \;\; i=j 

\end{array} \right. .$$

 Then, if $(S_{j})_{-1\le j \le d}$ is the sequence of symmetric
 sub-resultants computed with $S_{-1}=F$ and $S_{0}=G$, we have, for any $k=1,...
 , d$ :

$$ S_{k}(0)=(-1)^{k}.  f_{0} ^{k}.f_{d}^k \det({\mathcal{T}}_{k}(F,G)). $$

\end{proposition}
\begin{proof}{Proof : } As we have  $G=(-u-\sum_{i>0}u_{i}X^{-i})F$,
the following sequence of relations holds : \begin{eqnarray*} g_{0} & = &
-uf_{0}-u_{1}f_{1}- \cdots -u_{d-1}f_{d-1} - u_{d}f_{d}, \\
g_{1} & = & -uf_{1}-u_{1}f_{2}- \cdots -u_{d-1}f_{d}, \\ &
\vdots &  \\ g_{d} & = &  -uf_{d}.  \end{eqnarray*}
                
Now define for $k=1, ... , d$, the following three $k\times k$ matrices : $$
\widetilde{\mathbf{F}}_{k}=\left( \begin{array}{cccc}

f_{d}& 0 &  & 0 \\

f_{d-1}& f_{d} &  &  \\

\vdots &  & \ddots &  \\

f_{d-k+1}& \cdots & \cdots & f_{d} 

\end{array} \right), \;\;
\widetilde{\mathbf{G}}_{k}=\left( \begin{array}{cccc}

g_{d}& 0 &  & 0 \\

g_{d-1}& g_{d} &  &  \\

\vdots &  & \ddots &  \\

g_{d-k+1}& \cdots & \cdots & g_{d} 

\end{array} \right), $$

$$
\mathbf{U}_{k}=\left( \begin{array}{cccc}

u& 0 &  & 0 \\

u_{1}& u &  &  \\

\vdots &  & \ddots &  \\

u_{k-1}& \cdots & \cdots & u 

\end{array} \right) .  $$

Our relations can be translated by the following matricial relation : $$
\widetilde{\mathbf{G}}_{k}= -\mathbf{U}_{k}\cdot \widetilde{\mathbf{F}}_{k}. $$

Likewise, comparing the coefficients of $G=(v+\sum_{i>0}v_{i}X^{i}) F$,
we obtain : 

$$ \mathbf{G}_{k}=\mathbf{V}_{k}\cdot \mathbf{F}_{k} $$ with 

$$
\mathbf{F}_{k}=\left( \begin{array}{cccc}

f_{0}& \cdots & \cdots  & f_{k-1} \\

& f_{0} &  & f_{k-2}  \\

&  & \ddots & \vdots  \\

&  &  & f_{0} 

\end{array} \right), \;\; 
\mathbf{G}_{k}=\left( \begin{array}{cccc}

g_{0}& \cdots & \cdots  & g_{k-1} \\

& g_{0} &  & g_{k-2}  \\

&  & \ddots & \vdots  \\

&  &  & g_{0} 

\end{array} \right), $$
 
 $$ 
\mathbf{V}_{k}=\left( \begin{array}{cccc}

v& v_{1} & \cdots & v_{k-1}\\

 & v &  & v_{k-2}  \\

  &  & \ddots & \vdots  \\

 &  &  & v 

\end{array} \right).  $$
 
 These relations imply : $$ \left( \begin{array}{cc} \mathbf{I}_{k} &
\mathbf{I}_{k} \\ \mathbf{V}_{k} & - \mathbf{U}_{k} \end{array} \right)
\cdot \left( \begin{array}{cc} \mathbf{F}_{k} & \mathbf{0} \\
\mathbf{0} &  \widetilde{\mathbf{F}}_{k} \end{array} \right) =
\left( \begin{array}{cc} \mathbf{F}_{k} &
\widetilde{\mathbf{F}}_{k}\\ \mathbf{G}_{k} &
\widetilde{\mathbf{G}}_{k} \end{array} \right). $$ 

($\mathbf{I}_{k}$ denotes the identity matrix of order $k$.) Now, we can compute
the determinant of each side. For the left most matrix we subtract the
$i$-th column from the $(i+k)$-th one ($i=1,..., k$).  The result follows.
\end{proof}

\subsection{Signature of an Hermitian \textsc{Toeplitz} matrix}
Given an Hermitian \textsc{Toeplitz} matrix :

 $$ {\mathcal{ T }}_{d}=\left(
 \begin{array}{cccc} t_{0} & \bar{t}_{1}&\cdots & \bar{ t }_{d-1}\\ t_{1} &
\ddots &  & \vdots \\ \vdots & & \ddots &\bar{t}_{1} \\ t_{d-1}&
\cdots& t_{1} & t_{0}
 \end{array} \right),$$

we want to compute the signature of the associated Hermitian form. We didn't find any reference in the
literature to this simple problem, although there are several methods proposed
in the case of real \textsc{Hankel} matrices (see \cite{GE0} and \cite{SL}).

The signature of ${\mathcal{ T }}_{d}$ can be computed from the sequence of
 signs of its principal minors. The rule given by \textsc{Iohvidov}
{\cite{IO}} and, independently, by one of us \cite{SP}, works even when some
of these minors vanish. Once the minors are computed, the signature is obtained in
${{\mathcal{ O }}(d)}$ arithmetic operations.

The problem is then reduced to  the computation of the sequence of principal
minors of ${\mathcal{ T }}_{d}$. This can be achieved by the computation of the
constant terms of the sequence of the symmetric subresultants of two
polynomials as the next proposition shows.
 
\begin{proposition} Let ${{\mathcal{ T }}_{d}}$ be a Hermitian
\textsc{Toeplitz} matrix, defined as above, and $T$ the polynomial
: $$ T= -\bar{ t }-\bar{ t }_{1}X- \cdots - \bar{ t }_{d-1}X^{d-1}+
t_{d-1}X^d+ \cdots + t_{1}X^{2d-2}-tX^{2d-1}, $$ with $t\not= 0$ and
$t_{0}=t+\bar{t}$.

Let $(S_{j})_{-1 \le j \le 2d}$ be the sequence of symmetric subresultants
of $X^{2d-1}+1$ and $T$. For $j=1, ... , d$, we have : $$
\delta_{j}=S_{j}(0), $$ where $\delta_{j}$ is the $j$-th principal minor
of ${{\mathcal{ T }}_{d}}$.  \end{proposition}

\begin{proof}{Proof : } We can use Proposition \ref{linkssrtop} in this special
case. But the result can also be seen directly as well. Indeed, we have for
each $j=1, \ldots , d$ :

\begin{eqnarray*} S_{j}(0) & = & 
\underbrace{ \left | 
\begin{array}{ccc} 1 & &
\\ & \ddots & \\ && 1\\
     & & 0 \\ -\bar{ t } & \cdots & -\bar{ t }_{j-2} \\ & \ddots & \vdots \\ & &
     -\bar{ t } \\ & & 0 \end{array}
     \right. }_{j-1}
     \begin{array}{c} 0 \\ \vdots \\
\vdots \\ 1 \\ -\bar{ t }_{j-1} \\ \vdots \\ \vdots \\ -\bar{ t }
     \end{array} 
     \underbrace{\left. \begin{array}{cccc} 1 & & & \\ & \ddots & & \\ & &
\ddots & \\ & & & 1 \\ t & & & \\ \vdots & \ddots & & \\ \vdots &
& \ddots & \\ t_{j-1} & \cdots & \cdots & t \end{array} \right|}_{j}
\\ 
& = & \left| \begin{array}{ccccccccc} 1 &  &  & 0 &
& 0 & \cdots &\cdots &  0 \\ &\ddots & &  &  & \vdots &  &
& \vdots \\ & & \ddots &  & & \vdots & & & \vdots \\ 0 &  &
& 1 &       & 0 & \cdots& &  0 \\ -\bar{ t } & -\bar{ t }_{1}
& \cdots & -\bar{ t }_{j-1} & & t_{0} & \bar{ t }_{1} &
\cdots & \bar{ t }_{j-1}\\ & \ddots & \ddots & \vdots
&  & t_{1} & \ddots & & \vdots  \\ &  & \ddots & \bar{ t
}_{1}                                     &  & \vdots &
\ddots & \ddots & \bar{ t }_{1} \\ &  & & -\bar{ t }
&  & t_{j-1} & \cdots & t_{1} & t_{0} \end{array}
\right| \\ \\ & = & \delta_{j}.  \end{eqnarray*}
\end{proof}

Using \textbf{FSSR} Algorithm, we can then compute the signature of a Hermitian
\textsc{Toeplitz} matrix of order $d$ in ${{\mathcal{ O }}(d\log (d) ^2  \log \log (d))}$ arithmetical operations.

\textsc{Brunie} in \cite{BR} has shown that it is possible to improve
the algorithm  also to get the rank of the matrix, but this extra computation has
an arithmetical cost of ${{\mathcal{ O }}(d^2)}$ operations. There still
exists no fast solution to the rank problem.

\subsection{\textsc{Toeplitz} linear systems}

We now consider a much more popular application than the signature
problem.  Let ${{\mathcal{ T }}_{d}}$ be a \textsc{Toeplitz} matrix of
dimension $d$.  Suppose it is invertible and we want to compute
${{\mathcal{ T }}_{d}^{- 1}}$.  Several authors have given fast
algorithms to solve the problem.  \textsc{Brent}, \textsc{Gustavson}
and \textsc{Yun} in \cite{BGY} have a solution using \textsc{Pad\'e}
approximants, continued fractions and Euclidean algorithms.  Their
solution has a cost of ${{\mathcal{ O }}}(d\log (d) ^2 \log \log (d))$
arithmetical operations and uses the \textsc{Gohberg-Semencul}
formulae.  More recently \textsc{Gemigniani} in \cite{GE1} and
\cite{GE2} has used the \textsc{Schur} decomposition of a matrix with
the advantage that in defective cases no extra computation is needed.
Both algorithms have the same cost.  \textsc{Bini} and \textsc{Pan}
give in \cite{BP} the state of the art on this problem.

The solution developed here also works with the formulae of
\textsc{Gohberg-Semen\-cul}.  However we use the symmetric
subresultants; therefore we are able to manage the defective cases
directly with the \textbf{FSSR} algorithm without extra computation.
Our cost is the same as in \cite{BGY}, although, in defective cases, we
approximately divide computation time of by a factor two.
Furthermore, our algorithm is fraction free, until the last step.

As it is one of our tools, we recall first the  \textsc{Gohberg-Semencul}
formulae \cite{GS}.

\begin{theorem}\label{formGS} Let ${\mathcal{ T }}_{d}=(t_{i-j})_{0\le i,j \le
d-1}$ be an  invertible \textsc{Toeplitz} matrix. We denote by
$\mathbf{x}=(x_{0},\ldots , x_{d-1})^{t}$ the first column and by
$\mathbf{y}=(y_{0}, ... , y_{d-1})^{t}$ the last column of ${\mathcal{
T}}_{d}^{-1}$.  If $x_{0} \not=0$, we have :

\begin{eqnarray*} {\mathcal{ T }}_{d}^{-1} & =& \frac{1}{x_{0}} \left[
\left( \begin{array}{cccc} x_{0}& 0 & \cdots & 0 \\ \vdots & \ddots &
\ddots & \vdots \\ \vdots &   & \ddots & 0 \\ x_{d-1}& \cdots &
\cdots & x_{0} \end{array} \right) \cdot
\left ( \begin{array}{cccc} y_{d-1}& \cdots & \cdots & y_{0} \\ 0& \ddots &  &
\vdots \\ \vdots& \ddots  & \ddots  & \vdots \\ 0& \cdots & 0 & y_{d-1}
\end{array} \right) \right. \\ &  &\qquad \qquad \qquad  -
\left.  \left( \begin{array}{cccc} 0 & \cdots & \cdots & 0 \\

y_{0}& \ddots &  & \vdots \\

\vdots& \ddots & \ddots & \vdots \\

y_{d-2}& \cdots & y_{0} & 0 

\end{array} \right) \cdot
\left( \begin{array}{cccc}

0& x_{d-1} & \cdots & x_{1} \\

\vdots& \ddots & \ddots & \vdots \\

\vdots &  & \ddots & x_{d-1} \\

0 & \cdots & \cdots & 0 

\end{array} \right) \right]. \;\;\;\;(*) \end{eqnarray*}

If $x_{0}=0$, there exists an extension ${\mathcal{ T }}_{d+1}=(t_{i-j})_{0\le
i,j \le d}$ of ${\mathcal{ T }}_{d}$ which is invertible and such that the first
column of ${\mathcal{ T }}_{d+1}^{-1}$, say $\mathbf{\tilde{x}}=(\tilde{x}_{0},
\ldots ,\tilde{x}_{d})$, has its first coordinate different from zero. Let
$\mathbf{\tilde{y}}=(\tilde{y}_{0}, \ldots ,\tilde{y}_{d})$ denote the last column of
${\mathcal{ T }}_{d+1}^{-1}$. In this case, we have : \begin{eqnarray*}
 {\mathcal{ T }}_{d}^{-1} & =& \frac{1}{\tilde{x}_{0}} \left[
\left( \begin{array}{cccc} \tilde{x}_{0}& 0 & \cdots & 0 \\ \vdots &
\ddots & \ddots & \vdots \\ \vdots &   & \ddots & 0 \\
\tilde{x}_{d-1}& \cdots & \cdots & \tilde{x}_{0} \end{array}
\right) \cdot
\left ( \begin{array}{cccc} \tilde{y} _{d}& \cdots & \cdots & \tilde{y }_{1} \\
0& \ddots &  & \vdots \\ \vdots& \ddots  & \ddots  & \vdots \\ 0& \cdots
& 0 & \tilde{y} _{d} \end{array} \right) \right. \\ & & \qquad \qquad \qquad - 
\left.  \left( \begin{array}{cccc} \tilde{y }_{0} & 0 & \cdots & 0 \\

\vdots & \ddots & \ddots  & \vdots \\

\vdots&  & \ddots & 0 \\

\tilde{y }_{d-1}& \cdots & \cdots &  \tilde{y }_{0}  

\end{array} \right) \cdot
\left( \begin{array}{cccc}

 \tilde{x}_{d} & \cdots & \cdots & \tilde{x}_{1} \\

0 & \ddots &  & \vdots \\

\vdots & \ddots & \ddots & \vdots \\

0 & \cdots & 0 & \tilde{x}_{d} 

\end{array} \right) \right].\;\;\;\; (**) \end{eqnarray*} \end{theorem}

Therefore, if ${\mathcal{ T }}_{d}=(t_{i-j})_{0\le i,j \le d-1}$ is an
invertible \textsc{Toeplitz} matrix, the problem is reduced to the computation
of the vectors $\mathbf{x}$ and $\mathbf{y}$ or $\mathbf{\tilde{x}}$ and
$\mathbf{\tilde{y}}$ depending on the situation. We can use the symmetric
subresultants algorithm for this task.

Let us define the two polynomials : $$ S_{-1}=X^{2d+1}+1, $$ $$ S_{0}=
T_{\gamma, \delta}= - t_{-}-t_{-1}X- \cdots - t_{-d+1}X^{d-1} + \gamma X^d+ \delta
X^{d+1}+ t_{d-1} X^{d+2}+ \cdots + t_{+} X^{2d+1}, $$

where coefficients $t_{+}$ and  $t_{-}$ are different from 0 and satisfy
$t_{+}+t_{-}=t_{0}$. The complex coefficients $\gamma$ and $\delta$ will be
determined later on during the computation in order to apply Theorem
\ref{formGS}.

One can note that from $F=S_{-1}$ and $G=S_{0}$ we can rebuild the
matrix $T$ using Proposition 11 : we have $T= T_{d}(S_{-1}, S_{0})$.

Let $(S_{j})_{-1\le j \le 2d+1}$ be the sequence of symmetric
subresultants computed with $S_{-1}$ and $S_{0}$. 
As ${\mathcal{ T }}_{d}$ is
invertible, we have $S_{d}(0)=(-1)^d \det ({\mathcal{ T }}_{d})\not=0$ (use
Proposition \ref{linkssrtop}).
We will write
$S_{d}=\sum_{i=0}^{d+1}s_{i}X^{i}$. There also exist two polynomials $U_{d-1}=
\sum_{i=0}^{d-1}u_{i}X^{i}$ and $V_{d-1}=\sum_{i=0}^{d-1}v_{i}X^{i}$,
such that : $$ X^{d-1}S_{d} =
U_{d-1}(X^{2d+1}+1) + V_{d-1} T_{\gamma, \delta}. $$

This relation can be translated into  matricial terms as follows : 

$$
\left( \begin{array}{cccc} 1 & 0  & \cdots &  0 \\ 0 & \ddots & \ddots & \vdots
\\ \vdots & \ddots & \ddots & 0 \\ 0 & \cdots & 0 & 1 \\ \hline 0 &
\cdots & \cdots & 0 \\ \vdots &  &  & \vdots \\ & & & \\ \vdots &  &  &
\vdots \\ 0 & \cdots & \cdots & 0 \\ \hline 1 & 0  & \cdots &  0 \\ 0 &
\ddots & \ddots & \vdots \\ \vdots & \ddots & \ddots & 0 \\ 0 & \cdots &
0 & 1 \end{array} \right) \cdot \left( \begin{array}{c} u_{0}\\ \vdots
\\ \vdots \\ u_{d-1} \end{array} \right)+ $$

$$
 \left( \begin{array}{cccc}

-t_{-}  & 0  & \cdots &  0 \\

- t _{-1} & \ddots & \ddots & \vdots \\

\vdots & \ddots & \ddots & 0 \\

- t _{-d+1} & \cdots & - t _{-1} & - t _{-} \\ \hline -\gamma & -t_{-d+1} &
\cdots  & - t _{-1} \\

\delta & \ddots & \ddots & \vdots \\

\vdots & \ddots & \ddots & \vdots \\

\vdots  & \ddots & \ddots  & - \gamma  \\

t_{1} &  & t_{d-1} & \delta \\ \hline t_{+} & t_{1}  & \cdots &  t_{d-1} \\

0 & \ddots & \ddots & \vdots \\

\vdots & \ddots & \ddots & t_{1} \\

0 & \cdots & 0 & t_{+}

\end{array} \right) \cdot \left( \begin{array}{c}

v_{0}\\

\vdots \\

\vdots \\

v_{d-1}

\end{array} \right) =
\left( \begin{array}{c}

0 \\

\vdots\\

0 \\

s_{0} \\

\hline s_{1} \\

\vdots \\

\vdots \\

\vdots \\

s_{d+1} \\

\hline 0 \\

\vdots \\

\vdots \\

0

\end{array} \right)
\begin{array}{cc}

{\left.  \matrix{\cr \cr \cr \cr \cr} \right\}} & d \\

\left.  \matrix{\cr \cr \cr \cr \cr \cr \cr} \right\} & d+1\\

\left.  \matrix{\cr \cr \cr \cr \cr} \right\} & d 

\end{array} .  $$

If we subtract the first $d$ lines from the last $d$ ones, we obtain :
$$ {\mathcal{ T }_{d}}^t \left( \begin{array}{c} v_{0} \\

\vdots \\

\vdots \\

v_{d-1}

\end{array} \right)= \left( \begin{array}{c} 0\\

\vdots \\

0 \\

-s_{0}

\end{array}\right), $$

with $s_{0}=S_{d}(0)=(-1)^{d}\det ({\mathcal{ T }}_{d})\not=0$. Therefore, we
see that $\displaystyle \frac{-1}{s_{0}}\left( \begin{array}{c} v_{d-1}\\

\vdots \\

\vdots \\

v_{0}

\end{array} \right)$ is the first column of  ${\mathcal{ T
}}_{d}^{-1}$.  The same trick applied to ${\mathcal{ T
}}_{d}^{t}$ gives the last column of our matrix.
If $v_{d-1} \not=0$, we can apply the
first formula of \textsc{Gohberg-Semencul} to conclude.

By the proof of Lemma \ref{relbezout},  we get $v_{d-1}=-S_{d-1}(0)=(-1)^d \det ({{ \mathcal{ T }}}_{d-1}
)$. If $v_{d-1}=0$, we have to
compute the next symmetric subresultants, $S_{d+1}$. There exist two polynomials, $U_{d}$
and $V_{d}$, of degree at most $d$, such that : $$
X^dS_{d+1}=U_{d}(X^{2d+1}+1)+V_{d}T_{\gamma, \delta}. $$

In this case,  $\deg(V_{d})=d$, because
$\mathrm{co}_{d}(V_{d})=v_d=(-1)^{d+1}\det({\mathcal{
T }}_{d})\not=0$. If $S_{d+1}(0)\not=0$, we see, by the same computation as in the generic case just
above, that the coefficients of $-V_{d}/S_{d+1}(0)$ determine the first column of
the inverse of :

$$ {\mathcal{ T }}_{d+1}=\left( \begin{array}{c|c}

{\mathcal{ T }}_{d} & \begin{array}{c}

\gamma \\

 t_{-d+1}\\

\vdots \\ t_{-1}

\end{array}  \\ \hline \begin{array}{cccc}

\delta & t_{d-1} & \cdots & t_{1} 

\end{array}& t_{0} 

\end{array} \right). $$

Therefore we have to choose the coefficients $\gamma$ and $\delta$ in
 order to satisfy $S_{d+1}(0)=(-1)^d \det({\mathcal{ T }}_{d+1}) \not = 0$.

\begin{proposition} Using the above definitions,  suppose that
$\det({\mathcal{ T }}_{d-1})=0$ and $\det({\mathcal{ T }}_{d})\not=0$.
Define the three vectors of dimension $d$ : $$ \mathbf{V_{-}}=
\left( \begin{array}{c} 0\\

t_{-d+1}\\

\vdots \\

t_{-1}

\end{array} \right), \mathbf{V_{+}}= \left( \begin{array}{c} 0\\

t_{d-1}\\

\vdots \\

t_{1}

\end{array} \right) \;\; \mathrm{and}\;\; \mathbf{e_{0}}= \left(
\begin{array}{c} 1\\

0\\

\vdots \\

0

\end{array} \right).  $$

Then, the determinant of ${\mathcal{ T }}_{d+1}$ satisfies : 

$$ \det({\mathcal{ T
}}_{d+1})=-\det({\mathcal{ T }}_{d})\cdot( \gamma
\mathbf{V_{+}}^t\mathcal{T}_{d}^{-1}\mathbf{e}_{0} + \delta \mathbf{e}_{0}^t
\mathcal{T}_{d}^{-1}
\mathbf{V_{-}}+\mathbf{V_{+}}^t\mathcal{T}_{d}^{-1}\mathbf{V_{-}}-t_0 ). 
$$ 

Furthermore, in the above relation, the coefficients
$\mathbf{V_{+}}^t\mathcal{T}_{d}^{-1}\mathbf{e}_{0}$ and
$\mathbf{e}_{0}^t \mathcal{T}_{d}^{-1} \mathbf{V_{-}}$, of $\gamma$
and $\delta$ respectively, cannot vanish.
\end{proposition}

\begin{proof}{Proof :} We can factorize ${\mathcal{ T }}_{d+1}$ as follows : $$
{\mathcal{ T }}_{d+1}=
\left( \begin{array}{c|c}

{\mathcal{ T }}_{d} & \begin{array}{c}

0 \\

\vdots \\

0

\end{array}  \\ \hline \begin{array}{cccc}

\delta & t_{d-1} & \cdots & t_{1}

\end{array} & f \end{array} \right) \cdot
\left( \begin{array}{c|c}

{\mathcal{ I }}_{d} &

\begin{array}{c}

 \cr

r\\

\cr

\end{array}  \\ \hline \begin{array}{ccc}

0 &  \cdots & 0

\end{array}& 1 

\end{array} \right), $$ with $r={\mathcal{ T }}_{d}^{-1}\left( \begin{array}{c}

 \gamma \\

t_{-d+1}\\

\vdots\\

t_{-1}

\end{array} \right)= {\mathcal{ T }}_{d}^{-1}(\gamma
\mathbf{e_{0}}+\mathbf{V_{-}})$ and : 
$$ f= t_{0}-(\delta \mathbf{e_{0}} +\mathbf{V_{+}})  ^t \cdot
{\mathcal{ T }}_{d}^{-1}(\gamma \mathbf{e_{0}}+\mathbf{V_{-}}).$$

Then, we  have : $$ f=t_{0}-(\gamma \delta \cdot \mathbf{e_{0}}^t {\mathcal{ T
}}_{d}^{-1}\mathbf{e_{0}}+ \gamma \cdot \mathbf{V_{+}}^t {\mathcal{ T
}}_{d}^{-1}\mathbf{e_{0}}+ \delta \cdot \mathbf{e_{0}}^t{\mathcal{
T}}_{d}^{-1}\mathbf{V_{-}} + \mathbf{V_{+}}^t {\mathcal{ T
}}_{d}^{-1}\mathbf{V_{-}}).$$ 

But $\mathbf{e_{0}}^t {\mathcal{ T }}_{d}^{-1}\mathbf{e_{0}}$ is, up to the
factor $1/\det({\mathcal{T}}_d)$, equal to $\det({\mathcal{ T }}_{d-1})$ which is zero. 
Therefore, we obtain the stated formula.

We know that ${\mathcal{ T }}_{d}$ is invertible; let $(x_{0}, \ldots , x_{d-1})^t$
be the first column of its inverse.  Since $\det({\mathcal{ T }}_{d-1})=0$, we
have $x_{0}=0$. If we suppose that \linebreak $\mathbf{V_{+}}^t {\mathcal{ T
}}_{d}^{-1}\mathrm{e_{0}}=0$, we have $(0, t_{d-1},\ldots , t_{1})\cdot \left(
\begin{array}{c}

0\\

x_{1}\\

\vdots\\

x_{d-1}

\end{array} \right)=0$, and we can write~:

$$ \left( \begin{array}{c|c}

{\mathcal{ T }}_{d} & \begin{array}{c}

0 \\ t_{-d+1}\\

\vdots \\

t_{-1}

\end{array}  \\ \hline \begin{array}{cccc}

0 & t_{d-1} & \cdots & t_{1}

\end{array} & t_{0} \end{array} \right) \cdot \left( \begin{array}{c}

0\\

x_{1}\\

\vdots\\

x_{d-1}\\ \hline 0 \end{array} \right) = \left( \begin{array}{c}

1\\

0\\

\vdots\\

0\\ \hline 0 \end{array} \right) $$ $$ = \left( \begin{array}{c|c}

t_{0} & 

\begin{array}{cccc}

t_{-1} &  \cdots & t_{-d+1} & 0

\end{array}\\ \hline

\begin{array}{c}

t_{1} \\

\vdots \\

t_{d-1}\\ 0

\end{array} &
  
{\mathcal{ T }}_{d} 

\end{array} \right) \cdot  \left( \begin{array}{c}

0\\ \hline x_{1}\\

\vdots\\

x_{d-1}\\

0 \end{array} \right) $$

We therefore conclude that : $$ {\mathcal{ T }}_{d}\cdot \left( \begin{array}{c}
x_{1}\\

\vdots\\

x_{d-1}\\
0 \end{array} \right)=0.$$ 

However, as ${\mathcal{ T }}_{d}$ is invertible, the
equation ${\mathcal{ T }}_{d}\cdot X=0$ has only one solution, that is the zero vector.
 This leads to a contradiction since $x_{1},\ldots , x_{d-1}$ are not all equal
to zero. Therefore, the coefficient $\mathbf{V_{+}}^t {\mathcal{
T}}_{d}^{-1}\mathbf{e_{0}}$ cannot vanish. A similar argument works with
$\mathbf{e_{0}}^t{\mathcal{ T}}_{d}^{-1}\mathbf{V_{-}}$.  \end{proof}

Now we are able to choose a pair $(\gamma, \delta)$ such that
$\det({\mathcal{ T }}_{d+1})\not=0$.  In fact, as the set of pairs
$(\gamma, \delta)$ that make $\det({\mathcal{ T }}_{d+1})$ zero is a
line, after three attempts we are guaranteed to find an acceptable
value (for example, we try $(0, 0)$, then $(0,1)$ and if, with both
values, the determinant is zero, we can then use $(1,0)$ as a good
coefficient).

Before we describe the algorithm for fast inversion of a \textsc{Toeplitz}
matrix, we have to make some important remarks.

First, the polynomials $U_{d-1}$ and $V_{d-1}$ defined by:

$$X^{d-1}S_{d} =
U_{d-1}(X^{2d-1}+1) + V_{d-1} T_{\gamma, \delta}, \;\;\; (\ddagger)$$

are obtained  from \textbf{FSSR} applied to $X^{2d+1}+1$ and $T_{\gamma,\delta}$
with $r=d+2$.
As $S_{d}(0)\not=0$, if $\deg S_d=d+1$, there exists $k_\ell$ such that $k_{\ell}=d$.
We can then compute $M_{d}$.  The coefficients on the second
line of this matrix, $M_{d}$, are exactly $U_{d-1}$ and $V_{d-1}$, as we can see
from the proof of Proposition \ref{prop:transitmatrices}.  

Otherwise, if $\deg S_d < d+1$, we observe that for the biggest $\ell$ such that $k_\ell<d$ 
we have the pair $(S_{k_\ell}, S_{k_\ell}+1)$ right-defective (indeed
 Theorem \ref{grostheo} shows that all other situations lead to $S_k(0)=0$ for $k_\ell<k<k_{\ell+1}$). We know that in this case $S_{k_\ell} +1$ and $S_d$ are proportional ; the coefficient of proportionality is given by Theorem \ref{grostheo}. From \textbf{FSSR} we obtain only :

$$X^{k_{\ell}}S_{k_\ell+1} =
U_{k_{\ell}}(X^{2d-1}+1) + V_{k_{\ell}} T_{\gamma, \delta},$$

Multiplication by the right coefficient provides formula $(\ddagger)$.

Furthermore, whatever the situation might be, in this
call to \textbf{FSSR}, $\gamma$ and $\delta$ do not occur because we use a
truncation to the order $d-1$.

This provides the first column of ${\mathcal{ T }}_{d}^{-1}$. The same computation applied to $X^{2d-1}+1$
and $\bar{S}_{0}^{*}$ gives the last column.

Next, we do not need any extra call to \textbf{FSSR} when we test,  for example,
$(\gamma,\delta)=(0,0)$, $(1,0)$ or $(0,1)$. The computations are different only for
the last step, the transition from $S_{d}$ to $S_{d+1}$, and we do not need
to begin again the computation from $S_{-1}$ and $S_{0}$.  This is the first
advantage of our \textbf{FITM} algorithm over the one in \cite{BGY}. 
A second advantage  is that it is fraction-free.

\begin{figure}[th] {\small \begin{tabular}{lp{10cm} } \hline \multicolumn{2}{c}
{ALGORITHM \textbf{FITM}}\rule[-3pt]{0pt}{18pt} \\ \hline \hline
\textbf{INPUT :} \rule[-3pt]{0pt}{18pt}& ${\mathcal{ T
}}_{d}=(t_{i-j})_{0\le i,j \le d-1}$, a \textsc{Toeplitz} matrix of
dimension $d$ \\

\textbf{OUTPUT :} & ${\mathcal{ T }}_{d}^{-1}$ if ${\mathcal{ T }}_{d}$ is
invertible and, if not, a message that ${\mathcal{ T }}_{d}$ is
not invertible \\

\hline\\

\textbf{INITIALISATION }\rule[-3pt]{0pt}{18pt} &-- $S_{-1}=X^{2d+1}+1$ \\ &--
$S_{0}=T_{0,0}= -  t - t _{-1}X- \cdots -  t _{-d+1}X^{d-1} + t_{d-1} X^{d+2}+
\cdots + t X^{2d+1}, $\\

\textbf{MAIN PART :} & -- $ \mathbf{FSSR}(S_{-1}, S_{0}, d+2)$ \\ &\hspace{2cm} \%
we get $M_{k_{l}}$ with \\ &\hspace{2cm} \% $k_{l}$
the largest index such that $k_{l} \le d$.\\
 &   -- if $k_{l}=d$ and $S_{k_{l}}(0)=0$ or if $k_{l}< d$, and $S_{k_{l}+1}(0)=0$,\\
&\hspace{2cm} ${\mathcal{ T
}}_{d}$ is not invertible. \textbf{STOP} \\ 
& --  compute $V_{d-1}$ from $M_{k_{l}}$ and possible use of Theorem 3 \\

& -- $ \mathbf{FSSR}(S_{-1}, \bar{S}_{0}^{*}, d+2)$ \\ &\hspace{2cm} \% we get
$\widetilde{S}_{\widetilde{k_{l}}}, \widetilde{U}_{k_{l}-1}, \widetilde{V}_{k_{l}-1}$
with \\ &\hspace{2cm} \% $\widetilde{k_{l}}$ the largest index such that
$\widetilde{k_{l}} \le d$.\\ 
& -- compute $\widetilde {V}_{d-1}$ from $\widetilde{M_{k_{l}}}$  and possible use of Theorem 3 \\

& -- If $\deg V_{d-1}=d-1$, then ${\mathcal{ T }}_{d}^{-1}$ is computed via 
formula  (*)\\ 
& -- If $\deg \widetilde{V}_{d-1}=d-1$, then $({\mathcal{ T
}}_{d}^{t})^{-1}$ is computed via  formula \hspace*{1ex} (*)\\ 
& -- If $\deg V_{d-1}<d-1$ and $\deg \widetilde{V}_{d-1}<d-1$,
compute $S_{d+1}$ \hspace*{1ex} using $M_{k_{l}}$.\\ 
& -- If
$S_{d+1}(0)\not=0$, then ${\mathcal{ T }}_{d}^{-1}$ is computed via  formula
(**)\\ & -- otherwise redo the computation of $S_{d+1}$ with $T_{0,1}$ or with
\hspace*{1ex} $T_{1,0}$.\\ &\hspace{2cm} \% one of them will give $S_{d+1}(0)\not=0$.\\

\textbf{END.}& \\ \hline \end{tabular} } \end{figure}

Finally we can rewrite  our result in a \textsc{Toeplitz}-Bezoutian form. If $(U,V)$ is a pair of polynomials of degree at most $d$ such that 
$$X^d S_{d+1}(S_{-1},S_{0})=(X^{2d+1}+1)V+UP,$$
and if $(u,v)$ is a pair of polynomials of degree at most $d$ such that 
$$X^d S_{d+1}(S_{-1},S_{0}^{*})=(X^{2d+1}+1)v+uP,$$
then, in the non-degenerative situation, we have :
$$Bez(U^*,u) T_{d}(S_{-1}, S_{0})=S_{d}(0)S_{d+1}(0)I_{d}$$
where $I_{d}$ is the identity matrix of order $d$. (It comes from a well-known matrix representation  of Bezoutian - see \cite{BP}, p.156.)

There are certainly relations between our computations and those proposed by
\textsc{Gemigniani} in \cite{GE1} and \cite{GE2}. Bezoutians are used instead
of symmetric sub-resultants. But, these algorithms start with quite the same
polynomials.  In the literature one finds several links between resultants and 
Bezoutians (see for example \cite{KN}). However, in our particular case, the
relation between these two methods is not easy to describe and will be the
object of future work. 

Of course, all that we have said in this sub-section can be simplified in the case of
a Hermitian \textsc{Toeplitz} matrix.  It has been described in detail in
\cite{BR}.

We can now summarize our results in the \textbf{FITM} algorithm for fast inversion
of a \textsc{Toeplitz} matrix.

\section{Conclusion}

We have generalized the concepts introduced for the improvement of the
\textsc{Schur-Cohn} algorithm.  The sequence of sub-resultants defined
for a pair $(P, P^*)$ can now be computed for a general pair of
polynomials and the fast algorithm designed in the previous situation
has been extended.

The effectiveness of the algorithms presented has been studied in
\cite{BR} where they have been effectively programmed in TP language,
using the DFT. It has been shown that the bounds are effective and
that, for polynomials of degrees greater than 300 and coefficients
bounded by $2^{32}$, these algorithms are faster than their counterpart
programmed without DFT.

Of course, the fast version of the \textsc{Schur-Cohn} algorithm has
not changed, but we can present applications to \textsc{Toeplitz}
matrices which are new.  It would be an interesting study to compare
the different algorithms for the inversion of \textsc{Toeplitz}
matrices and to explore the links between them.


\end{document}